\documentclass[apj]{emulateapj}

\usepackage{dcolumn}
\usepackage{amsmath,amssymb,amsthm,paralist}
\usepackage{graphicx}
\usepackage{color}
\usepackage{appendix}
\usepackage{subfigure}
\usepackage{longtable}
\usepackage{textcomp}
\usepackage{url}
\usepackage{verbatim}
\usepackage{lineno}
\usepackage{bm}
\usepackage{cancel}

\begin{document}


\title{Influence of Nuclear Reaction Rate Uncertainties on Neutron
Star Properties Extracted From X-ray
Burst Model-Observation Comparisons}



\author{Zach~Meisel, Grant~Merz, and Sophia~Medvid}
\email[]{meisel@ohio.edu}
\affiliation{Institute of Nuclear \& Particle Physics, Department of Physics \& Astronomy, Ohio University, Athens, Ohio 45701, USA}


\begin{abstract}
Type-I X-ray bursts can be used to determine properties of accreting
neutron stars
via comparisons between model calculations and astronomical
observations, exploiting the sensitivity of models
to astrophysical conditions. However, the sensitivity of models to
nuclear physics uncertainties calls into question the fidelity of
constraints derived in this way. Using X-ray burst model
calculations performed with the code {\tt MESA}, we investigate
the impact of uncertainties for nuclear reaction rates previously identified as
influential and compare them to the impact of changes in
astrophysical conditions, using the conditions that are thought
to best reproduce the source GS 1826-24 as a baseline. We find that
reaction rate uncertainties are unlikely to significantly change
conclusions about the properties of accretion onto the neutron star
surface for this source. However, we
find that reaction rate uncertainties significantly hinder the
possibility of extracting the neutron star mass-radius ratio by
matching the modeled and observed light curves due to the influence
of reaction rates on the modeled light curve shape. Particularly
influential nuclear reaction rates are $^{15}{\rm O}(\alpha,\gamma)$
and $^{23}{\rm Al}(p,\gamma)$, though other notable impacts arise
from $^{14}{\rm O}(\alpha,p)$, $^{18}{\rm Ne}(\alpha,p)$, $^{22}{\rm
Mg}(\alpha,p)$, $^{24}{\rm Mg}(\alpha,\gamma)$, $^{59}{\rm
Cu}(p,\gamma)$, 
and $^{61}{\rm Ga}(p,\gamma)$. Furthermore, we find that varying
some nuclear reaction rates within their uncertainties has an impact
on the neutron star crust composition and thermal structure that is comparable to
relatively significant changes accretion conditions.

\end{abstract}


\maketitle

\section{Introduction}
Thermonuclear explosions powered by hydrogen and helium burning on
the surfaces of accreting neutron stars, known as Type-I X-ray
bursts, have been the subject of modeling efforts since even before
they were first observed~\citep{Meis18b}. Early efforts established a
sensitivity of model calculation results to underlying neutron star
properties, such as the mass and radius, and the details of accretion
rate and accreted composition~\citep{Fuji81,Ayas82}. It has also been
understood since the beginning that the thermonuclear origin of the
bursts lends model calculations to a sensitivity to the nuclear
physics input~\citep{Joss77,Wall81}. An outstanding question is to what
extent the sensitivity of model results to nuclear physics
uncertainties limits the ability of model-observation comparisons to
extract constraints on astrophysical conditions by exploiting the
sensitivity of model results to these conditions. 

In the past decade several works have undertaken model-observation
comparisons to constrain the astrophysical conditions for specific
X-ray bursting sources. This has been enabled by the construction of
catalogues containing particularly well-characterized sources that
make for good modeling targets, e.g. \citet{Gall08,Gall17}. Examples
include constraints for the accretion properties of
SAX J1808.4-3658~\citep{John18} and accretion-based
heating for an object like EXO 0748-676~\citep{Keek17}. 

The most
common target for model-observation comparisons has been GS 1826-24,
selected for its textbook bursting behavior~\citep{Bild00,Gall08}.
\citet{Gall04} used a simple ignition model to constrain the
accretion metallicity for this source by matching the trend in the
burst recurrence time for increasing accretion rate.
\citet{Hege07}
constrained the accretion rate and metallicity by matching the shape
of the observed light curve for one bursting epoch 
to results from {\tt KEPLER}~\citep{Weav78,Woos04} model calculations
performed with an assumed neutron star mass and radius, which
\citet{Gall17} repeated for updated observational data.
\citet{Zamf12} employed the best-fit model from \citet{Hege07} to
constrain the neutron star compactness by a matching the observed
light curve flux and, for a separate constraint, the light curve
shape. \citet{Meis18a} used {\tt
MESA}~\citep{Paxt11,Paxt13,Paxt15,Paxt18} to perform a simultaneous matching of the
recurrence time and light curve shape for multiple bursting epochs
using a grid of simulated astrophysical conditions to simultaneously
constrain the accretion rate and composition, accretion-based
heating, and neutron star compactness. Note that each of the aforementioned studies adopted a single set
of nuclear physics input, though calculation results are known to be
sensitive to nuclear physics input.

\citet{Pari08} and \citet{Pari09} employed
post-processing calculations to demonstrate the sensitivity of X-ray
burst nucleosynthesis to nuclear reaction rates and masses,
respectively. More recently, single-zone (i.e. zero-dimensional)
models with self-consistent
temperature and density evolution have explored the impact of
uncertainties in nuclear reaction rates and nuclear masses on X-ray
burst light curves~\citep{Cybu16,Scha17}. \citet{Cybu16} also
employed multizone (i.e. one-dimensional) calculations
with {\tt KEPLER} to follow-up on reaction rates that were
identified as important in the single-zone calculations. Note that
the impact on the burst light curve has been described in these
works either using visual comparisons or an integral of the absolute
difference between the light curves of two models.

To date, it has not been directly explored how
nuclear physics uncertainties influence astrophysical constraints
derived from model-observation comparisons for a specific source. We
investigate this in the present work, employing the model
calculations described in \citet{Meis18a} and focusing on conditions
similar to those required to reproduce the observed features of GS
1826-24. This is presented as follows. Section~\ref{sec:calcs}
discusses the production of model light curves.
Section~\ref{sec:rates} describes the nuclear reaction rates for
which the model sensitivity is being assessed. Section~\ref{sec:lc}
examines the influence of varying these reaction rates within their
present uncertainties on model light curve properties, while
Section~\ref{sec:ashes} investigates the influence on model
nucleosynthesis.
Section~\ref{sec:compare} details the method for matching the shape
of a modeled light curve to observational data to extract a
mass-radius ratio and demonstrates the strong sensitivity of these
results to certain nuclear physics uncertainties.
Section~\ref{sec:discuss} concludes with a summary of our
findings.

\section{Model Calculations}
\label{sec:calcs}
The model calculations used in this work are the same as in
\citet{Meis18a}. The pertinent details are recapitulated here for
convenience.

\subsection{Code Details and Microphysics}
The code {\tt MESA} version 9793 was used to calculate the X-ray
luminosity over time and burst ashes produced by nuclear burning in $\sim$1000~zones
constructed to resemble an accreting neutron star envelope.
Corrections include a post-Newtonian modification of the local
gravity for general relativistic effects and a
time-dependent mixing length theory~\citep{Heny65,Paxt11} for
convection. The time resolution and spatial resolution adapt
according to the {\tt MESA} controls {\tt varcontrol\_target=1d-3} and
{\tt mesh\_delta\_coeff=1.0}~\citep{Paxt13}. The $\sim$0.01~km thick
envelope has an inner boundary of neutron star mass $M_{\rm
NS}=1.4$~$M_{\odot}$ and radius $R_{\rm NS}=11.2$~km. The 304~isotope
network of \citet{Fisk08} was used with REACLIB~\citep{Cybu10}
version 2.2 as the baseline nuclear reaction rate library. The solar 
metal fraction of \citet{Grev98} was used to distribute the accreted
metals.

Astrophysical conditions which were varied between different model
calculations include the accretion rate $\dot{M}$, metallicity of
the accreted composition $Z$, accretion-based heating at the base of
the envelope $Q_{\rm b}$, and hydrogen mass fraction $X(H)$. The helium
mass fraction $Y$ was adjusted to enforce $X(H)+Y+Z=1$. 
$\dot{M}=0.05, 0.07, 0.08, 0.11, 0.15, 0.17~\dot{M}_{\rm{E}}$, where
$\dot{M}_{\rm{E}}=1.75\times10^{-8}M_{\odot}$/yr
is the Eddington accretion rate~\citep{Scha99}, were explored to
sample $\dot{M}$ for the observed GS 1826-24
epochs from 1998, 2000, and 2007~\citep{Gall08} and the observed
$\dot{M}$ scaled up by $\sim$2~\citep{Meis18a}. $Z=0.01, 0.02$ were used to 
investigate the solar $Z$ favored by previous investigations of
GS 1826-24~\citep{Gall04,Hege07}
and a slight reduction from that value. 
$Q_{\rm{b}}=0.1, 0.5, 1.0$~MeV/u were chosen to mimic shallow
heating thought to occur in the outer layers of accreting neutron
stars, with the lower-bound roughly corresponding to heat from
electron-captures induced by accretion~\citep{Gupt07,Meis16} and the
upper bound corresponding to a typical amount of heating inferred
from model-observation comparisons of neutron star crust
cooling~\citep{Turl15}. While larger $Q_{b}$ have been
inferred~\citep{Deib15}, we did not explore these since this can lead
to short waiting-time bursts~\citep{Keek17}, which GS 1826-24 does not exhibit.
In {\tt MESA} base heating is
achieved by fixing the luminosity of the base of the envelope, so
that the base luminosity depended on $Q_{\rm{b}}$ and $\dot{M}$ of
the model. $X(H)=0.50, 0.55, 0.60, 0.65, 0.70, 0.75$ were investigated
to sample a range of conditions from hydrogen-poor to hydrogen-rich
while still maintaining a sufficient amount of hydrogen to avoid
pure helium bursts.

\subsection{Light Curve Construction}
To mitigate the numerical noise that leads to some burst-to-burst
variability and occasional sharp features in an individual burst light curve,
{\tt MESA} light curves for a sequence of bursts calculated with one
set of astrophysical conditions were combined to an average light
curve, as is frequently done with observational data. The first
burst in a sequence was excluded, since the lack of ashes from prior
bursts leads to an atypically energetic burst~\citep{Woos04}. Bursts
in a sequence were mapped onto a uniform time grid with a linear
spline, smoothed by averaging the luminosity over a $\pm$1~s time
window, and aligned in time by defining a luminosity
threshold being crossed at time $t=0$.
The light curves corresponding to a sequence (typically
$10-20$~bursts) were combined to
achieve an average light curve and an uncertainty band, where fewer
bursts in a sequence will generally result in a larger band. See
\citet{Meis18a} for an example of this process.

\section{Nuclear Reaction Rates Varied}
\label{sec:rates}
Nuclear reaction rates were varied for calculations employing
the astrophysical conditions recently found to best reproduce
light curve features of GS 1826-24~\citep{Meis18a}: $\dot{M}=0.17~\dot{M}_{\rm E}$,
$Z=0.02$, $Q_{b}=0.1$~MeV/u, $X(H)=0.70$. Due to computational expense,
the set of reaction rates under investigation was limited to the 19
rate variations found to have a significant impact on X-ray burst light curves
calculated with {\tt KEPLER} for similar astrophysical
conditions~\citep{Cybu16}. Rate variation factors of
10 and 100 were adopted for reactions proceeding through high and
low level density regions in the compound nucleus, respectively,
following the approach of \citet{Cybu16}. The sources for other rate
variation factors are discussed below. We note that the rate
variation factors should be considered plausible but somewhat generous 
nuclear physics uncertainties, where the intention is that more
rigorous uncertainty evaluations could be pursued for reactions
found to be influential.

\begin{figure*}[t]
\begin{center}
\includegraphics[width=2.65\columnwidth,angle=90]{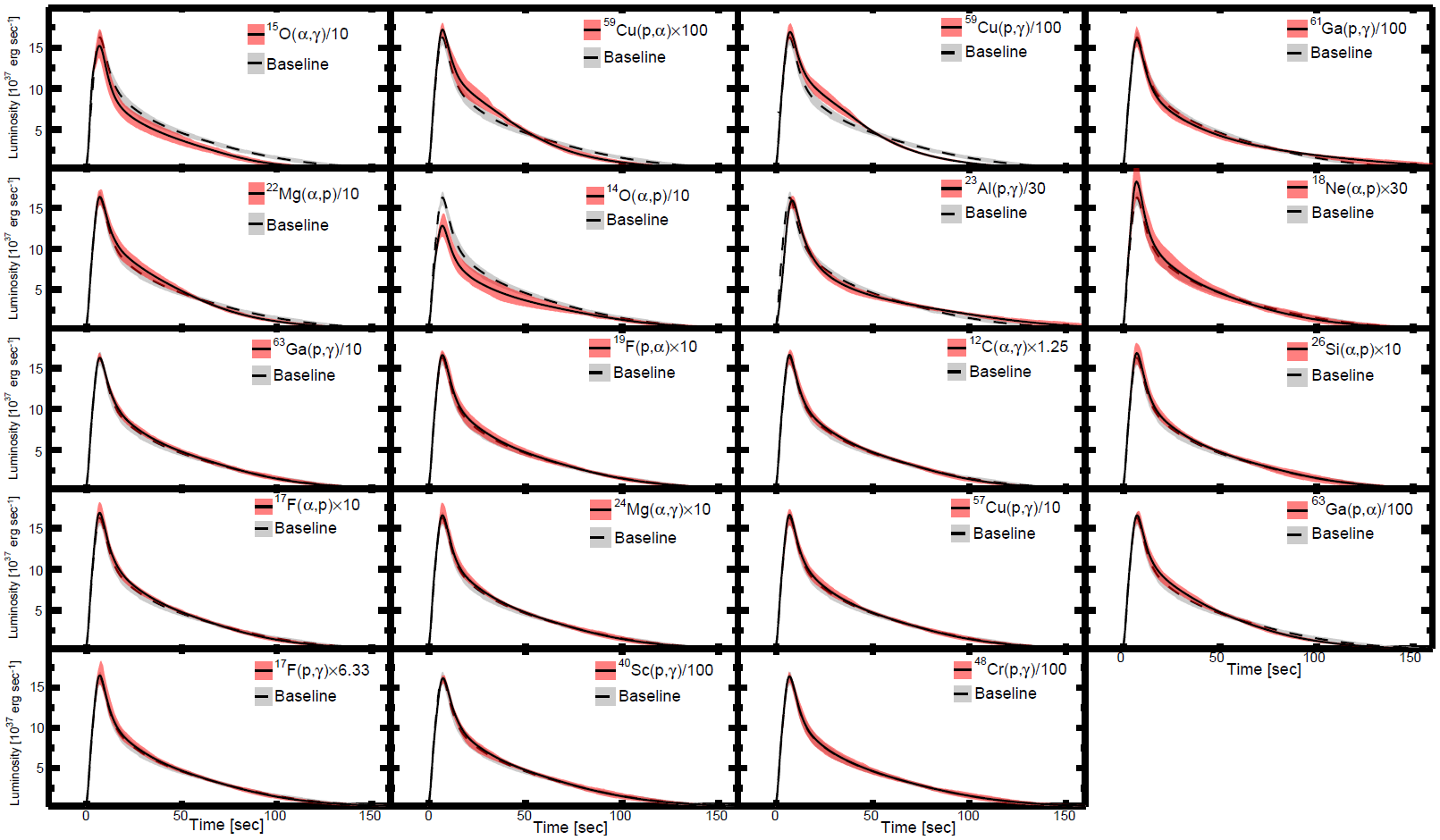}
\caption{Light curves calculated using a varied reaction rate (red
band) and a nominal reaction rate library (gray band) for the
best-fit astrophysical conditions of \citet{Meis18a}.
\label{fig:LCall}}
\end{center}
\end{figure*}

Detailed discussions of nuclear reactions relevant for X-ray bursts
can be found in
\citet{VanW94,Wies99,Woos04,Fisk04,Fisk08,Cybu16,Meis18b}.
The role of the rates varied in this work are briefly discussed
here in order to help
understand their potential to impact the X-ray burst light curve,
shown in Figure~\ref{fig:LCall}, and
resultant ash abundances. Listed in order of
multizone light curve impact according to \citet{Cybu16}:

\begin{enumerate}
\item $^{15}{\rm O}(\alpha,\gamma)^{19}{\rm Ne}$ was decreased by a
factor of 10 based on the lower-bound from the most recent
evaluation~\citep{Davi11}. This reaction is the location of breakout
from the hot CNO cycle, which is $\beta$-limited and thereby
concentrates abundances at $^{14}{\rm O}$ and $^{15}{\rm O}$, into
the $rp$-process.

\item $^{59}{\rm Cu}(p,\alpha)^{56}{\rm Ni}$\footnote{Note:
\citet{Cybu16} chose to write all positive $Q$-value $(p,\alpha)$
reactions as $(\alpha,p)$ reactions in their Table 2.} was increased
by a factor of 100. This reaction competes with $^{59}{\rm
Cu}(p,\gamma)$ in the NiCu cycle, working to keep material within
the cycle and prolonging the effective waiting-point at this
location.

\item $^{59}{\rm Cu}(p,\gamma)^{60}{\rm Zn}$ was decreased by a
factor of 100. This reaction is the avenue for breakout from the
aforementioned NiCu cycle.

\item $^{61}{\rm Ga}(p,\gamma)^{62}{\rm Ge}$ was decreased by a
factor of 100. This reaction connects the NiCu and ZnGa cycles by
proton-capture onto the $^{61}{\rm Ga}$ abundance maintained by the
$(p,\gamma)-(\gamma,p)$ equilibrium between $^{60}{\rm Zn}$ and
$^{61}{\rm Ga}$. This provides a bypass for the 
relatively long $^{60}{\rm Zn}$ half-life.

\item $^{22}{\rm Mg}(\alpha,p)^{25}{\rm Al}$ was decreased by a
factor of 10. This reaction occurs at a branch-point, competing with
the rather slow $^{22}{\rm Mg}$ $\beta^{+}$-decay and, at low temperatures, the
$^{22}{\rm Mg}(p,\gamma)$ reaction.

\item $^{14}{\rm O}(\alpha,p)^{17}{\rm F}$ was decreased by a factor
of 10 based roughly on the systematic change in the reaction rate
caused by assuming constructive or destructive interference between
resonances~\citep{Hu14} and on the fact that some disagreement
exists as to relevant resonance properties~\citep{Fort12}. 
This reaction connects hot CNO cycles, speeding up the
breakout process marking burst ignition.

\item $^{23}{\rm Al}(p,\gamma)^{24}{\rm Si}$ was decreased by a
factor of 30\footnote{While this work was in writing, we became
aware of an updated reaction rate uncertainty from C.~Langer et al. which
has been submitted for publication. The one standard
deviation uncertainty band for the new rate does not
include the factor of 30 reduction from the nominal reaction rate used
here.} based on the uncertainty evaluation of \citet{Cybu16} using
the $^{24}{\rm Si}$ structure information from \citet{Scha97}. This
reaction enables the $(p,\gamma)$ path from the $^{22}{\rm Mg}$
branch-point at high temperatures by proton-capture on the
$^{23}{\rm Al}$ present due to $(p,\gamma)-(\gamma,p)$
equilibrium.

\item $^{18}{\rm Ne}(\alpha,p)^{21}{\rm Na}$ was increased by a
factor of 30 based on the uncertainty evaluation of \citet{Cybu16}
using the experimental results of \citet{Mati09}. A smaller
uncertainty has been determined by a more recent
reaction rate evaluation~\citep{Mohr14}; however, a subsequent
evaluation of the $^{22}{\rm Mg}$ structure~\citep{Sham15} includes an additional
$1^{-}$ resonance in the Gamow window that would likely increase
the reaction rate if present. This reaction rate is a breakout
reaction from the hot CNO cycle due to the high $^{18}{\rm Ne}$
concentration that results from its $\beta$-limiting
role.

\item $^{63}{\rm Ga}(p,\gamma)^{64}{\rm Ge}$ was decreased by a
factor of 10. This reaction is the avenue for breakout from the
aforementioned ZnGa cycle.

\item $^{19}{\rm F}(p,\alpha)^{16}{\rm O}$ was increased by a factor
of 10. This reaction competes with the cold CNO cycle breakout
reaction $^{19}{\rm F}(p,\gamma)$ and the reverse rate is an
influential helium-burning reaction in burning regions lacking
hydrogen.

\item $^{12}{\rm C}(\alpha,\gamma)^{16}{\rm O}$ was increased by a factor of
1.25 to approximate the uncertainty band found by a recent
analysis of this reaction rate~\citep{Debo17}. This
reaction is a major source of energy generation where hydrogen
is absent, such as the reheated burst ashes below the region of burst
ignition.

\item $^{26}{\rm Si}(\alpha,p)^{29}{\rm P}$ was increased by a
factor of 10 based on the systematic uncertainty of \citet{Alma12}.
This reaction occurs at a branch-point, competing with
the $^{26}{\rm Si}(p,\gamma)$ reaction.

\item $^{17}{\rm F}(\alpha,p)^{20}{\rm Ne}$ was increased by a
factor of 10. This reaction competes with $^{17}{\rm F}(p,\gamma)$,
which bridges the CNO cycle into the hot CNO cycle.

\item $^{24}{\rm Mg}(\alpha,\gamma)^{28}{\rm Si}$ was increased by a
factor of 10. Like $^{12}{\rm C}(\alpha,\gamma)$, this reaction
generates energy in hydrogen-depleted zones during the
burst.

\item $^{57}{\rm Cu}(p,\gamma)^{58}{\rm Ni}$ was decreased by a
factor of 10\footnote{A rate reduction factor of 100 was used by
\citet{Cybu16} because their calculations were performed prior to
\citet{Lang14} reducing this rate uncertainty.}. This
reaction occurs on the equilibrium abundance of $^{57}{\rm Cu}$,
which is in $(p,\gamma)-(\gamma,p)$ equilibrium with $^{56}{\rm Ni}$
during most of the burst, expediting the flow out of this strong
waiting-point nucleus~\citep{Lang14}.

\item $^{63}{\rm Ga}(p,\alpha)^{60}{\rm Zn}$ was decreased by a
factor of 100. This reaction keeps material within the
aforementioned ZnGa cycle.

\item $^{17}{\rm F}(p,\gamma)^{18}{\rm Ne}$ was increased by a
factor of 6.33 based on the uncertainty estimate of \citet{Cybu16},
which was updated from the evaluation of \citet{Ilia10}. This reaction
connects the CNO and hot CNO cycles.

\item $^{40}{\rm Sc}(p,\gamma)^{41}{\rm Ti}$ was decreased by a
factor of 100. This reaction allows material to flow beyond the
$^{39}{\rm Ca}$ waiting-point, which is due to $Z=20$ magicity, by
proton-capture on the equilibrium abundance of $^{40}{\rm Sc}$.

\item $^{48}{\rm Cr}(p,\gamma)^{49}{\rm Mn}$ was decreased by a
factor of 100. This reaction pushes material beyond the $^{48}{\rm
Cr}$ bottleneck that occurs due to prior reaction pathways leading to this
relatively long-lived nucleus.
\end{enumerate}

\begin{figure*}[ht!]
  \centering
  \subfigure[Varying $\dot{M}$.]{\label{fig:LCm}\includegraphics[width=2in]{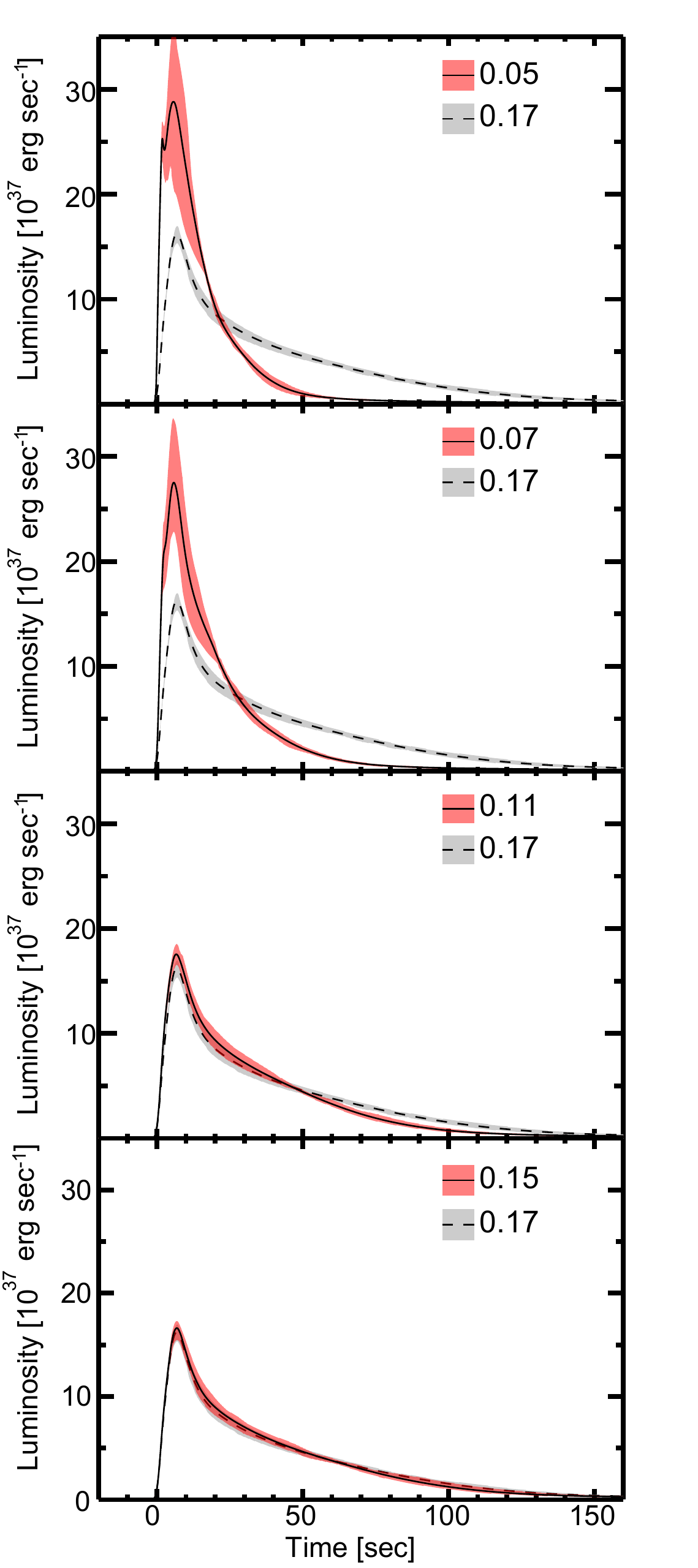}}
  \qquad
  \subfigure[Varying $X(H)$.]{\label{fig:LCh}\includegraphics[width=2in]{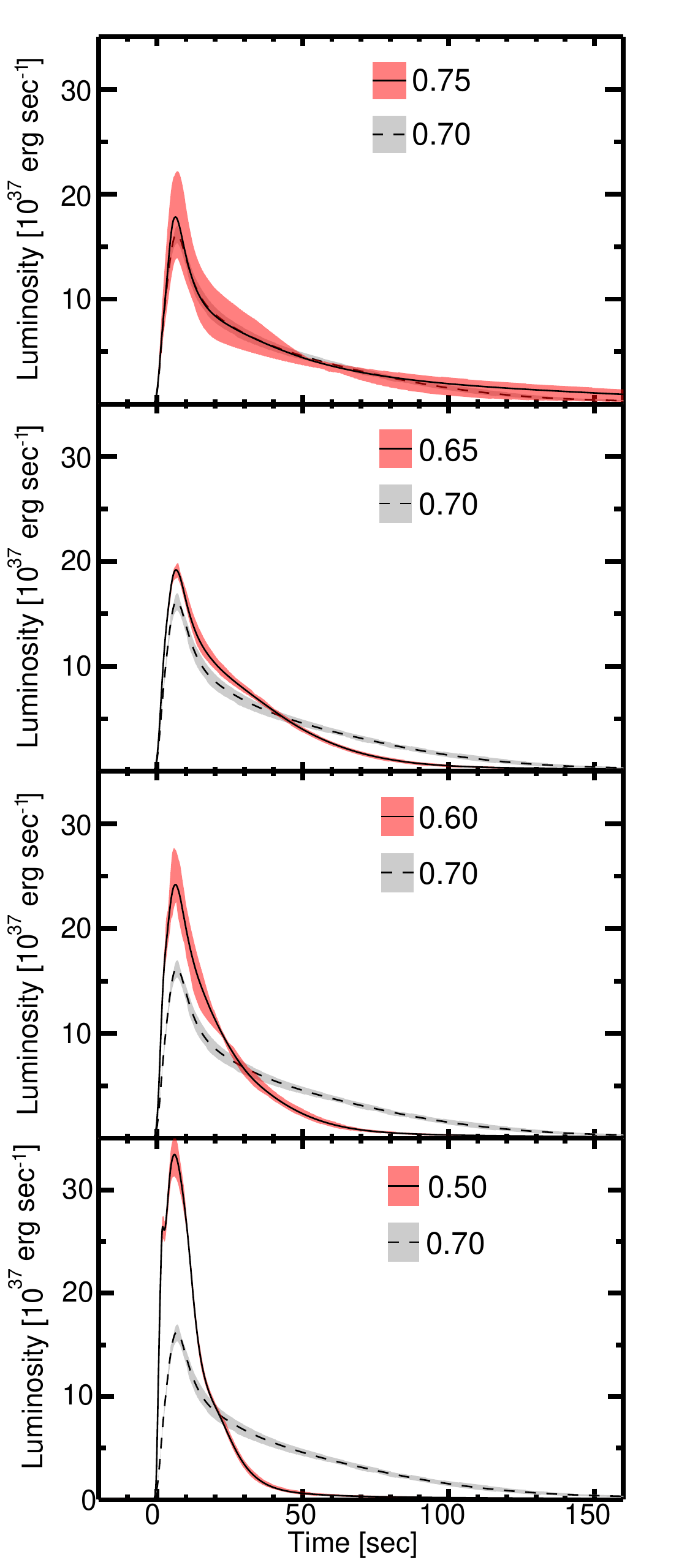}}
  \qquad
  \subfigure[Varying $Q_{\rm{b}}$ with $Z=0.01$.]{\label{fig:LCq}\includegraphics[width=2in]{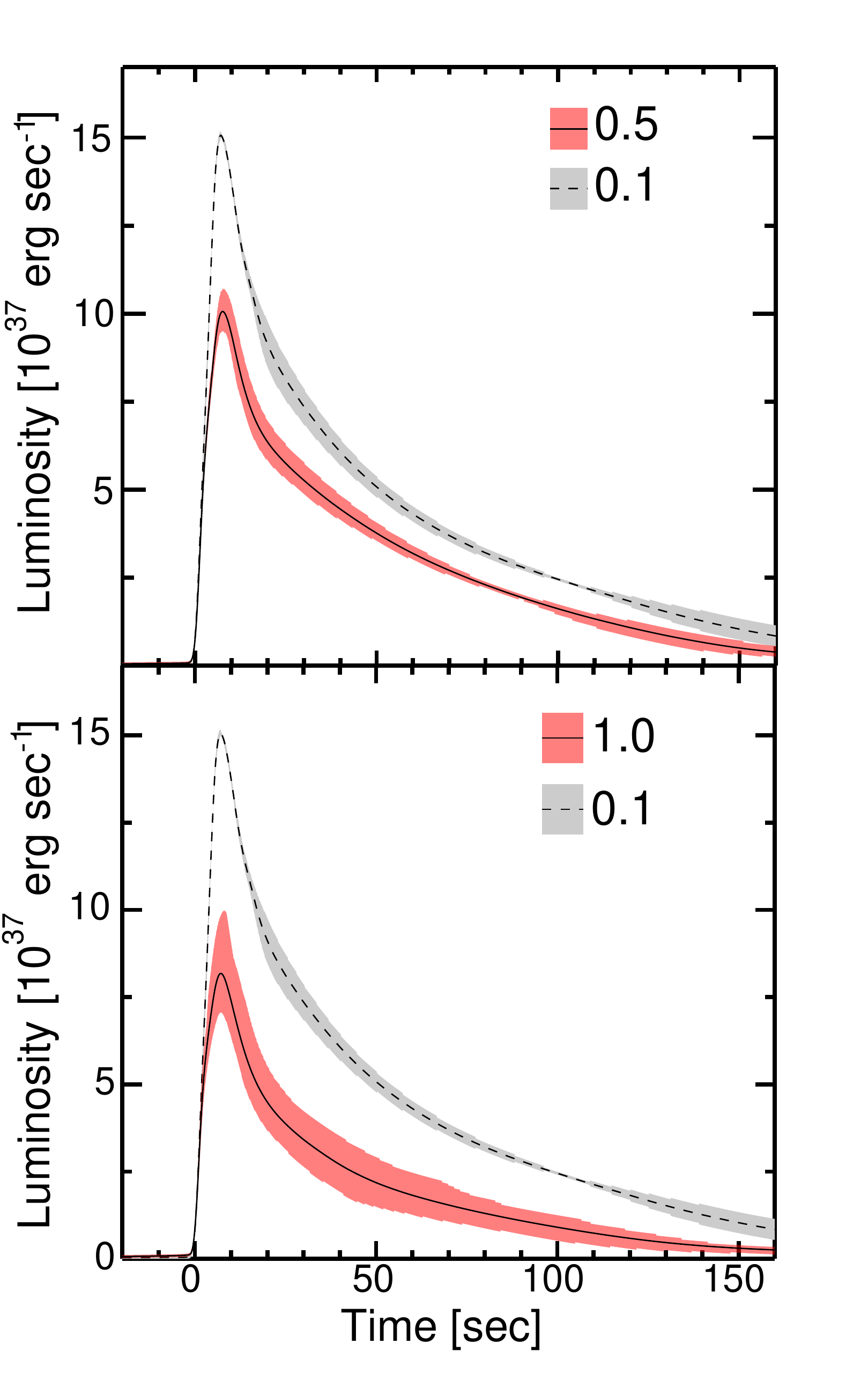}}
 \caption{Calculated light curves for, unless otherwise stated, the
 conditions
 $\dot{M}=0.17~\dot{M}_{\rm E}$, $X(H)=0.7$, $Z=0.02$, and $Q_{\rm
 b}=0.1$~MeV/u. The legends indicate the value of the quantity being
 varied in that figure panel.}
 \label{fig:LCastro}
\end{figure*}

\section{Influence on Light Curve Properties}
\label{sec:lc}

\subsection{Quantitative Light Curve Features}
The impact of astrophysical conditions and nuclear reaction rates on
the X-ray burst light curve was quantified by assessing their
influence on various light curve features. This
approach provides a more nuanced view than only quantifying the
overall deviation from one light curve to another using a
sensitivity factor, as done in \citet{Cybu16}. This also enables a
closer comparison to the X-ray burst calculations of \citet{Lamp16}, who
explored a large range of astrophysical conditions.
The light curve
features of interest here are the recurrence time $\Delta t_{\rm rec}$,
rise time $t_{\rm rise}$, ``10-to-10" time $t_{10-10}$,
convexity $\mathcal{C}$, non-exponentiality \cancel{E}, and the
Gaussian fluence fraction $f$, each of which are described below.

$\Delta t_{\rm rec}$ is the average time between X-ray bursts, where
the burst start is defined as the time when the luminosity passes
over a defined threshold.

$t_{10-10}$ is the time between the points in the X-ray
burst light curve rise and decay when the luminosity has risen above and
declined below 10\% of the peak luminosity for the averaged light
curve. This is a metric for the overall duration of the burst,
adopting the definition used in \citet{Cybu16}.

$\mathcal{C}$ was defined by \citet{Maur08} to quantify the shape of
the rise of the X-ray burst light curve. It is calculated by the
following procedure. First, the burst light curve rise is rescaled
so that the luminosity at 10\% of
the peak luminosity is at a rescaled (time,luminosity) of (0,0) and
the luminosity at 90\% of the peak luminosity is at (10,10).
$\mathcal{C}$ captures the deviation from linearity of the rise by
integrating the difference between the rescaled luminosity $L'$ and
a straight line: $\mathcal{C}=\int^{10}_{0}\left(L'-x\right)dx$.

We introduce \cancel{E} in analogy to $\mathcal{C}$, but for the
X-ray burst tail. To calculate this parameter, the burst light curve
is rescaled so that the peak luminosity has a rescaled
(time,luminosity) of (0,10) and the point in the X-ray
burst tail that the luminosity is $1/e$ of the peak luminosity is
defined as a (time,luminosity) of (10,$1/e$).
As the goal of \cancel{E} is to quantify the morphology of
the tail in order to capture the deviation from exponential
behavior, which has often been assumed for fitting
purposes~\citep{Zand17}, the integrated difference is mirrored at the
$1/e$ point:
\cancel{E}$=\int^{10}_{0}\left(L'-10e^{-x/10}\right)dx+\int_{10}^{50}\left(10e^{-x/10}-L'\right)dx$.

An alternative prescription for the light curve tail comes from
\citet{Zand17}. These authors fit an observed light curve flux using
a power-law to describe the cooling of the envelope and a Gaussian
to describe the extra flux due to $rp$-process burning:
$F(t)=F(t_{0})\left(t/t_{0}\right)^{-a}+\left(G/\sqrt{2\pi
s}\right)e^{-t^{2}/2s^{2}}$, where $F(t_{0})$, $a$, $G$, and $s$ are fit
to the data and $t_{0}$ is the time in the X-ray burst tail that the
flux first drops below 55\% of the peak flux. The $rp$-process contribution is
quantified by the fraction $f$ of the Gaussian fluence to the total
fluence. In order to stay within the same order of magnitude for $G$
as observed bursts, to calculate $f$ we apply a redshift $1+z=1.44$
and distance $d\xi_{\rm b}^{1/2}=6$~kpc to transform the simulated
averaged light curves into an ``observed" flux via 
$F=L/(4\pi\xi_{\rm{b}}(1+z)d^{2})$ and dilate time by a factor of
$1+z$.

Other than $f$, each of the above quantities was calculated for
individual bursts in a burst sequence and this information was used
to determine an average and 68\%
confidence interval.

\subsection{Impact of Astrophysical Conditions}
\begin{figure*}[ht!]
  \centering
  \subfigure[Varying
  $\dot{M}$.]{\label{fig:Im}\includegraphics[width=0.63\columnwidth]{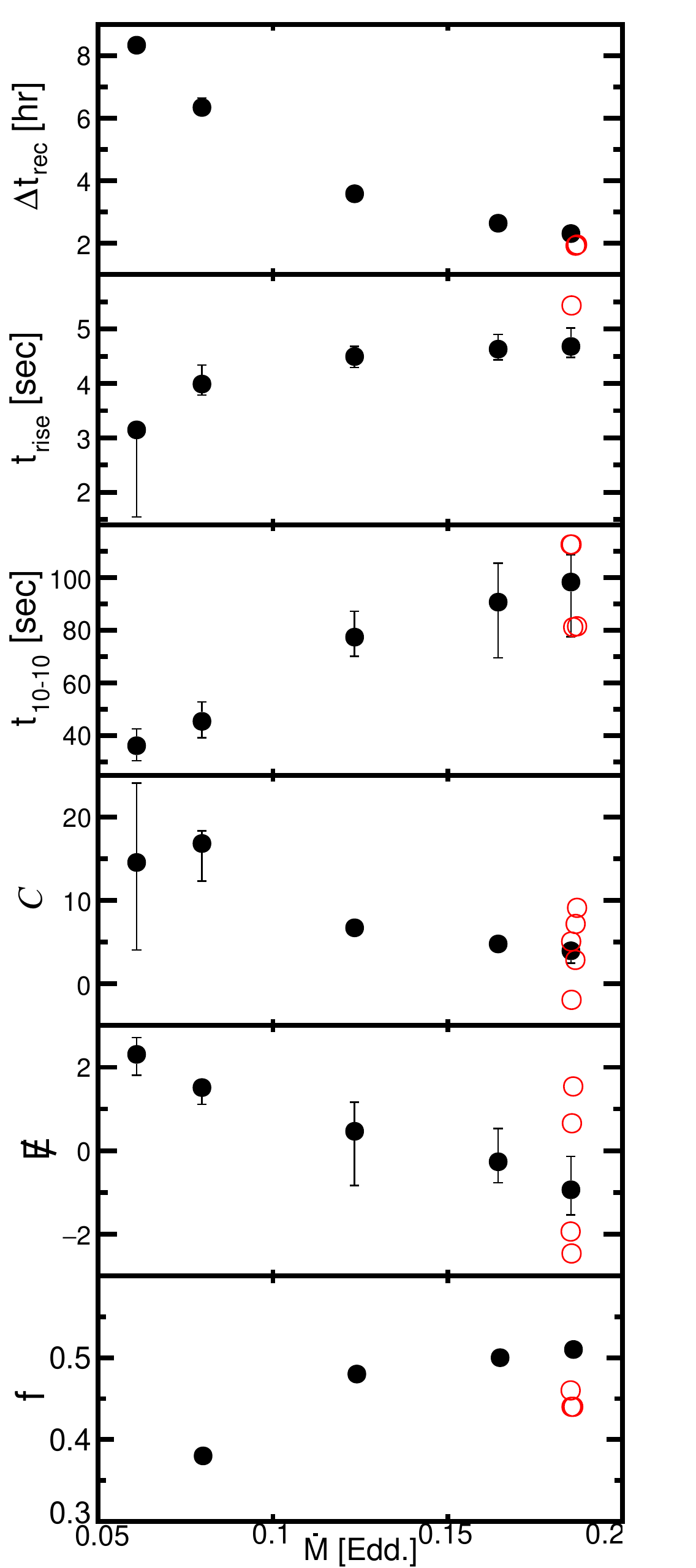}}
  \qquad
  \subfigure[Varying
  $X(H)$.]{\label{fig:Ih}\includegraphics[width=0.63\columnwidth]{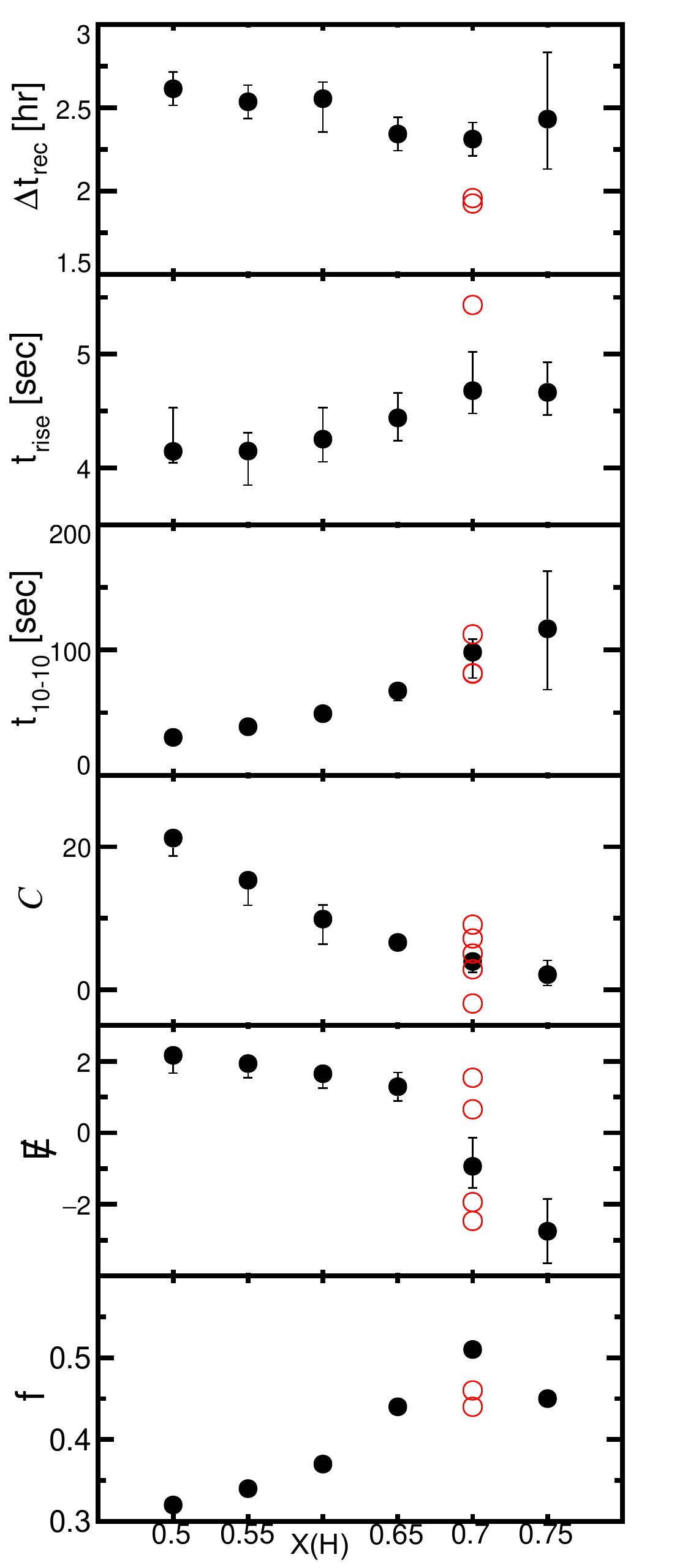}}
  \qquad
  \subfigure[Varying $Q_{\rm{b}}$ with
  $Z=0.01$.]{\label{fig:Iq}\includegraphics[width=0.63\columnwidth]{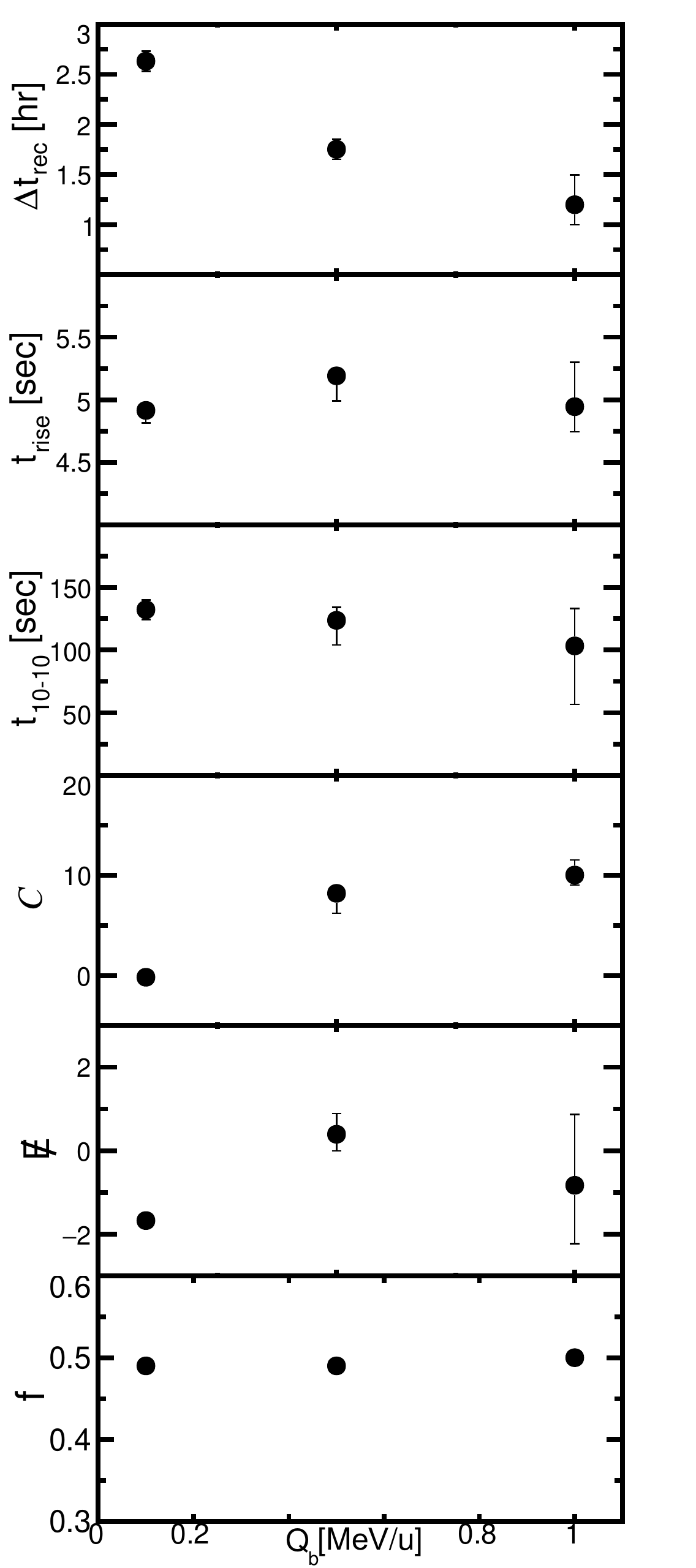}}
 \caption{Quantitative features for the light curves shown in
 Figure~\ref{fig:LCastro} (solid black circles) and cases from
 Figure~\ref{fig:LCall} that significantly deviated from the
 baseline result (open red circles), as discussed in
 Section~\ref{ssec:NucImpact}.}
 \label{fig:Compare}
\end{figure*}

The consequences for varying the accretion properties $\dot{M}$,
$X(H)$, and $Q_{\rm b}$ are shown in Figures~\ref{fig:LCastro} and
\ref{fig:Compare}. These effects can mostly be understood by
considering the way in which modifying an astrophysical condition
alters the H/He fraction and the envelope temperature in between
bursts, as these are the main two quantities impacting burst
ignition~\citep{Gall04}.

Qualitatively, it is
apparent that lower
$\dot{M}$, lower $X(H)$, and higher $Q_{\rm b}$, relative to the best-fit
calculations of \citet{Meis18a}, all shorten the duration of
the light curve. The former two modifications also increase the peak
luminosity of the burst, while the latter decreases the peak
luminosity. These changes can be understood by considering the
stable burning occurring between bursts. 

Decreasing $\dot{M}$
results in a cooler envelope since $Q_{\rm b}$ scales with accretion
rate. To compensate, a more He-rich environment is required to
achieve burst ignition conditions. 
$\Delta t_{\rm rec}$ lengthens for lower $\dot{M}$ since more
hydrogen must
be burned into helium to achieve the required He abundance. 
The more He-rich fuel will
burn brighter and faster, since helium burning reactions are generally
more temperature sensitive than hydrogen burning reactions. The faster
burning is manifested by reduced $t_{\rm rise}$ and $t_{10-10}$,
where the latter reduction is compounded by an absence of late time
hydrogen burning. Increased $\mathcal{C}$ for more He-rich burst
conditions are likely due to an emphasis of the $(\alpha,p)$-process
in the reaction flow~\citep{Wein06} and the flow favoring the slower
$(\alpha,p)$ reactions over $(p,\gamma)$ reactions at branch points
due to the lower hydrogen abundance at burst ignition.
\cancel{E} is positive for lower
$X(H)$, meaning the burst tail falls off faster than exponential,
indicating a lack of late-time hydrogen burning. The decrease in $f$
contains the same signature.

Directly reducing $X(H)$ makes for a more helium rich envelope,
leading to most of the same consequences as just discussed. However,
longer burning time between bursts is not required to build up the
helium density required for burst ignition, so $\Delta t_{\rm rec}$
is far less sensitive to changes in $X(H)$. Interestingly, for the
highest $X(H)$, $f$ begins to turn-over although there is 
more hydrogen burning ongoing at late times, i.e. a more extensive
$rp$-process.

Increasing $Q_{\rm b}$ has the effect of increasing the envelope
temperature, which reduces the burning time required to achieve
ignition, shortening $\Delta t_{\rm rec}$. Since less fuel is
built-up between bursts, the bursts will be less energetic,
exhibiting a lower peak luminosity, and have a shorter $t_{10-10}$.
We speculate that the higher temperatures modify the reaction flow
and therefore nuclear energy generation
during the burst rise. This would lead to increased $\mathcal{C}$ for
higher $Q_{\rm b}$. We leave the investigation of this phenomenon
for a future work, but note that it limits the ability for an
increased $Q_{\rm b}$ to allow for a decreased $\dot{M}$
while still explaining $\Delta t_{\rm rec}$ of an observed bursting
		  source. We discuss this further, as well as the
		  consequences, in Section~\ref{ssec:relative}.

The changes in the X-ray burst light curve properties with $\dot{M}$
are in good agreement with the behavior \citet{Lamp16} calculated
using {\tt KEPLER} models, with the exception of $\Delta t_{\rm
rec}$. While the same qualitative behavior is observed between the
two models, the {\tt KEPLER} models generally have a shorter $\Delta
t_{\rm rec}$ at a given $\dot{M}$ and a weaker dependence of
$\Delta t_{\rm rec}$ on $\dot{M}$. Resolving this discrepancy will
require a detailed comparison of model assumptions.

\subsection{Impact of Nuclear Reaction Rates}
\label{ssec:NucImpact}
Figures~\ref{fig:LCall} and \ref{fig:Compare} show the impact 
varied reaction rates have
on quantitative features of the X-ray burst light curve for the
baseline astrophysical conditions adopted here.
Note that the
results of all 19 rate variations are not shown in
Figure~\ref{fig:Compare}, rather only those
that have a significant impact on the quantity under consideration.

$\Delta t_{\rm rec}$ is reduced by 15\% for the rate
variations $^{15}{\rm O}(\alpha,\gamma)/10$ and $^{14}{\rm
O}(\alpha,p)/10$, which can be compared to the 11\% and 7\% respective
reductions found by \citet{Cybu16} for similar (but not identical) 
astrophysical conditions. 

$t_{\rm rise}$ is only significantly
altered by $^{23}{\rm Al}(p,\gamma)/30$, with an increase of 15\%.

$t_{10-10}$ is decreased by 20\% for $^{15}{\rm
O}(\alpha,\gamma)/10$ and $^{59}{\rm Cu}(p,\gamma)/100$ and increased
by 10\% for $^{61}{\rm Ga}(p,\gamma)/100$ and $^{23}{\rm
Al}(p,\gamma)/30$.

$\mathcal{C}$ is sensitive to several reaction rate variations,
ranging from 225\% of the baseline value for $^{15}{\rm
O}(\alpha,\gamma)/10$ to a 150\% reduction for $^{23}{\rm
Al}(p,\gamma)/30$. Other
notable sensitivities are an increase by 75\% for $^{14}{\rm
O}(\alpha,p)/10$, a decrease by 25\% for $^{18}{\rm
Ne}(\alpha,p)\times30$, and an increase by 25\% for $^{24}{\rm
Mg}(\alpha,\gamma)\times10$.

~\cancel{E} 
also shows a marked sensitivity to several rate
variations. 
The largest and second largest \cancel{E} correspond to $^{59}{\rm
Cu}(p,\gamma)/100$ and $^{22}{\rm Mg}(\alpha,p)/10$, respectively.
The most negative \cancel{E} corresponds to $^{23}{\rm
Al}(p,\gamma)/30$, while the
second most negative \cancel{E} is due to $^{61}{\rm
Ga}(p,\gamma)/100$.

$f$ is also altered by $^{23}{\rm Al}(p,\gamma)/30$, $^{59}{\rm
Cu}(p,\gamma)/100$, and $^{61}{\rm Ga}(p,\gamma)/100$; however, it
decreases for each rate variation. As such, though $f$ and
\cancel{E} describe the evolution of the burst light curve tail, it
appears they contain complementary information.

Overall, the
reaction rate sensitivities are qualitatively similar to those seen
for the multizone calculations of \citet{Cybu16}. However, that work did not report most of the
quantitative light curve features under study here and so a detailed comparison
is generally not possible. Furthermore, some differences could be
attributed to the somewhat lower accretion rate, $0.1~\dot{M}_{\rm
E}$, adopted in that work.

\subsection{Discussion of Relative Light Curve Impacts}
\label{ssec:relative}
Comparing the relative impacts of varied astrophysical conditions
and nuclear reaction rates on the features of the X-ray burst light
curve enables us to assess the robustness of constraints on
astrophysical conditions derived from model-observation comparisons,
e.g. from \citet{Hege07,Gall17,Meis18a}. It is important to note
that both the shape and $\Delta t_{\rm rec}$ must be reproduced and,
to limit the degeneracy of acceptable solutions, this should be done
for bursting epochs at multiple accretion rates if possible~\citep{Meis18a}.

From Figure~\ref{fig:Compare} it is apparent that $\Delta t_{\rm
rec}$ is most sensitive to changes in $\dot{M}$ and $Q_{\rm b}$.
Considering only $\Delta t_{\rm rec}$, one can see that the
($\dot{M}$,$Q_{\rm b}$) combinations (0.17~$\dot{M}_{\rm
E}$,0.1~MeV/u) and (0.11~$\dot{M}_{\rm E}$,1.0~MeV/u) would arrive
at a similar result. However, when including the burst morphology,
we see that this change in $\dot{M}$ leaves $\mathcal{C}$ largely
unchanged, but modifies \cancel{E} substantially. In opposition,
this change in $Q_{\rm b}$ significantly changes $\mathcal{C}$ and
leaves \cancel{E} more or less unaltered. One could imagine that the
$\mathcal{C}$ increase from increasing $Q_{\rm b}$ could be
mitigated by a modification of $X(H)$. However, it is apparent that
an $X(H)$ increase would only modestly reduce $\mathcal{C}$ while
also substantially reducing \cancel{E}. As such, to explain a source
such as GS 1826-24, there appears to be a unique solution in terms
of $\dot{M}$, $X(H)$, and $Q_{\rm b}$, which \citet{Meis18a} used 
to demonstrate that shallow heating in neutron star
outer layers can be constrained with X-ray burst light curve model-observation
comparisons.

It is natural to ask whether reaction rate variations could modify
these conclusions, resulting in a different set of best-fit
astrophysical conditions. For this analysis we turn to the unfilled
red circles of Figure~\ref{fig:Compare}. The first thing to notice
is that no rate variation substantially changes $\Delta t_{\rm
rec}$, so the two ($\dot{M}$,$Q_{\rm b}$) solutions explaining this
observable are robust. Then, to modify the conclusion about which of
these solutions is the best-fit, a reduction to $\mathcal{C}$ would
be required, \emph{while leaving other observables unchanged}. The
only candidate for this is the rate variation $^{23}{\rm Al}(p,\gamma)/30$.
However, this rate variation also increases $\Delta t_{\rm rise}$,
increases $t_{10-10}$, and decreases \cancel{E}. Therefore, it
appears that the constraints on the astrophysical conditions
involving accretion are robust. 

It is unlikely this conclusion would
be altered by incorporating the affects of flame spreading on the
neutron star surface, as this phenomenon will primarily impact only
the light curve rise~\citep{Maur08}. Similarly, while a modification of the
assumed surface gravitational redshift could change $\Delta t_{\rm
rise}$ and $t_{10-10}$ in the same direction (See
Section~\ref{sec:compare}), this light curve compression/stretching
would rescale the decay but not alter \cancel{E}.
However, it is possible that
different rate sensitivities for the higher $Q_{\rm b}$ conditions
may open-up alternative solutions to matching the observed GS
1826-24 light curve, so a definite conclusion would
require a nuclear reaction rate sensitivity study for those
conditions. Furthermore, as this analysis is limited to explaining
observed bursts for GS 1826-24, it cannot be certain that the
accretion conditions for other sources would be able to be
constrained in this way to the degree achieved here. Another caveat
is that extra mixing processes in the neutron star outer layers may
alter $\Delta t_{\rm rec}$ for a set of accretion conditions, though
work considering this phenomenon is still in the exploratory
phase~\citep{Cave17}.

While we find nuclear reaction rate variations do not modify the
conclusions about accretion properties for GS 1826-24 drawn from
model-observation comparisons, this does not mean that reaction rate
uncertainties are inconsequential for this source. In the following
section we show that they are of consequence for neutron star crust
properties. In Section~\ref{sec:compare}, we demonstrate that
nuclear reaction rate uncertainties have significant consequences
for mass-radius constraints which can be extracted from X-ray burst
model-observation comparisons.

\begin{figure*}[ht!]
  \centering
  \subfigure[Varying
  $\dot{M}$.]{\label{fig:Im}\includegraphics[width=0.60\columnwidth]{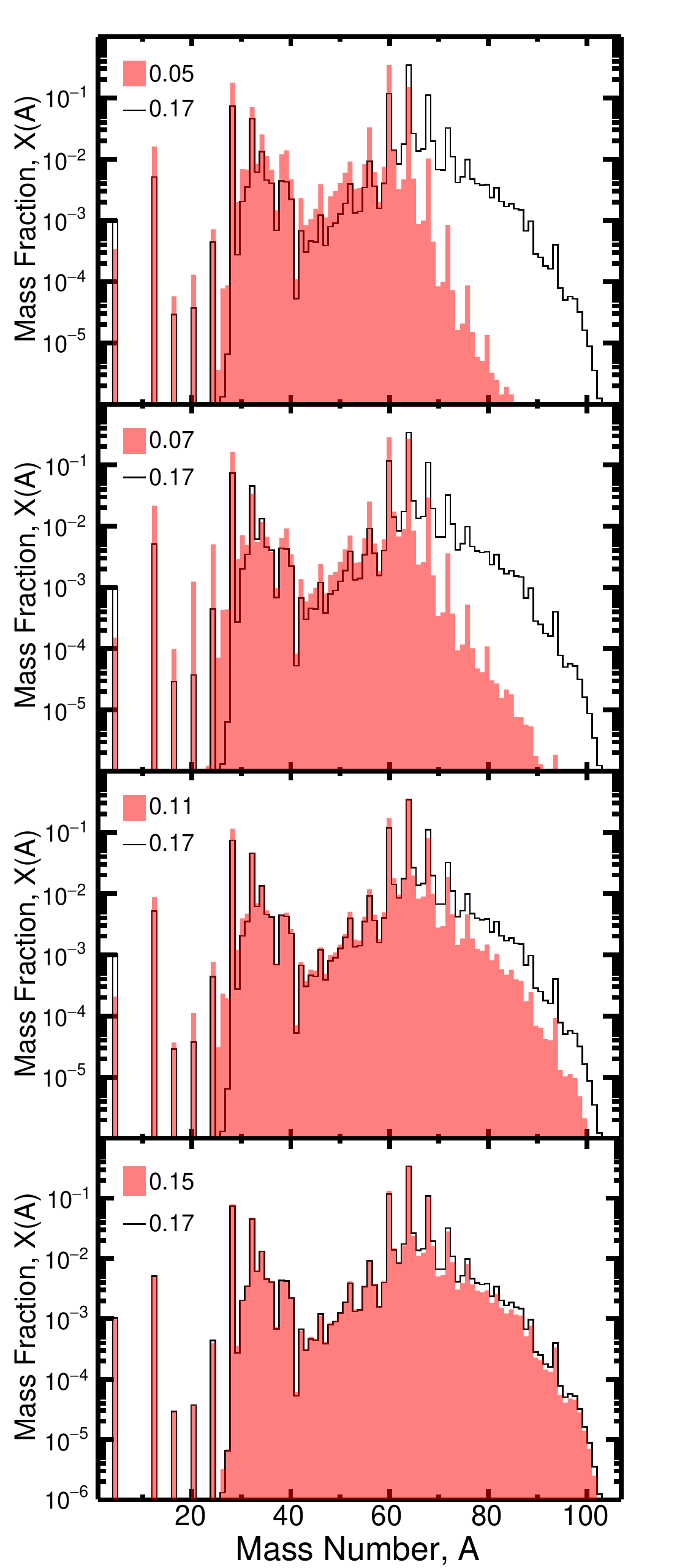}}
  \qquad
  \subfigure[Varying
  $X(H)$.]{\label{fig:Ih}\includegraphics[width=0.60\columnwidth]{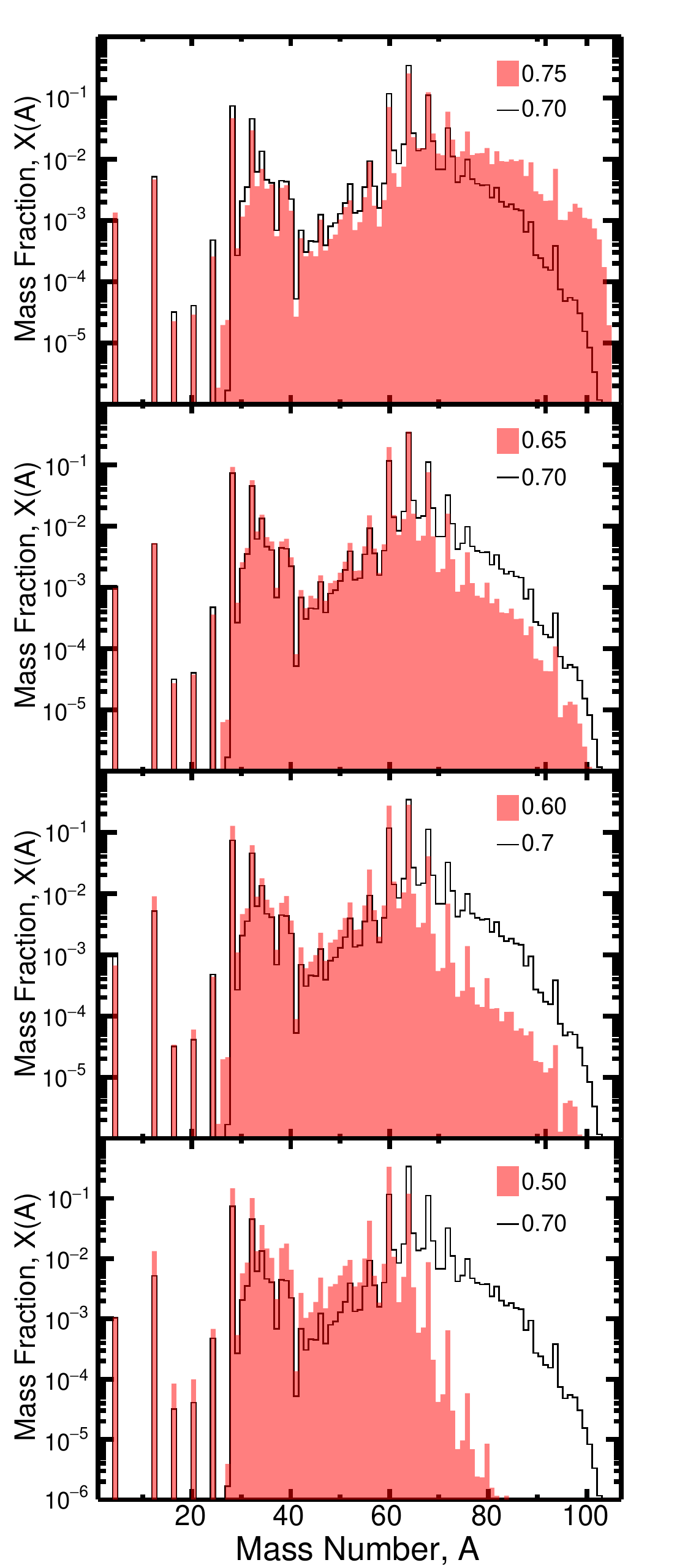}}
  \qquad
  \subfigure[Varying $Q_{\rm{b}}$ with
  $Z=0.01$.]{\label{fig:Iq}\includegraphics[width=0.60\columnwidth]{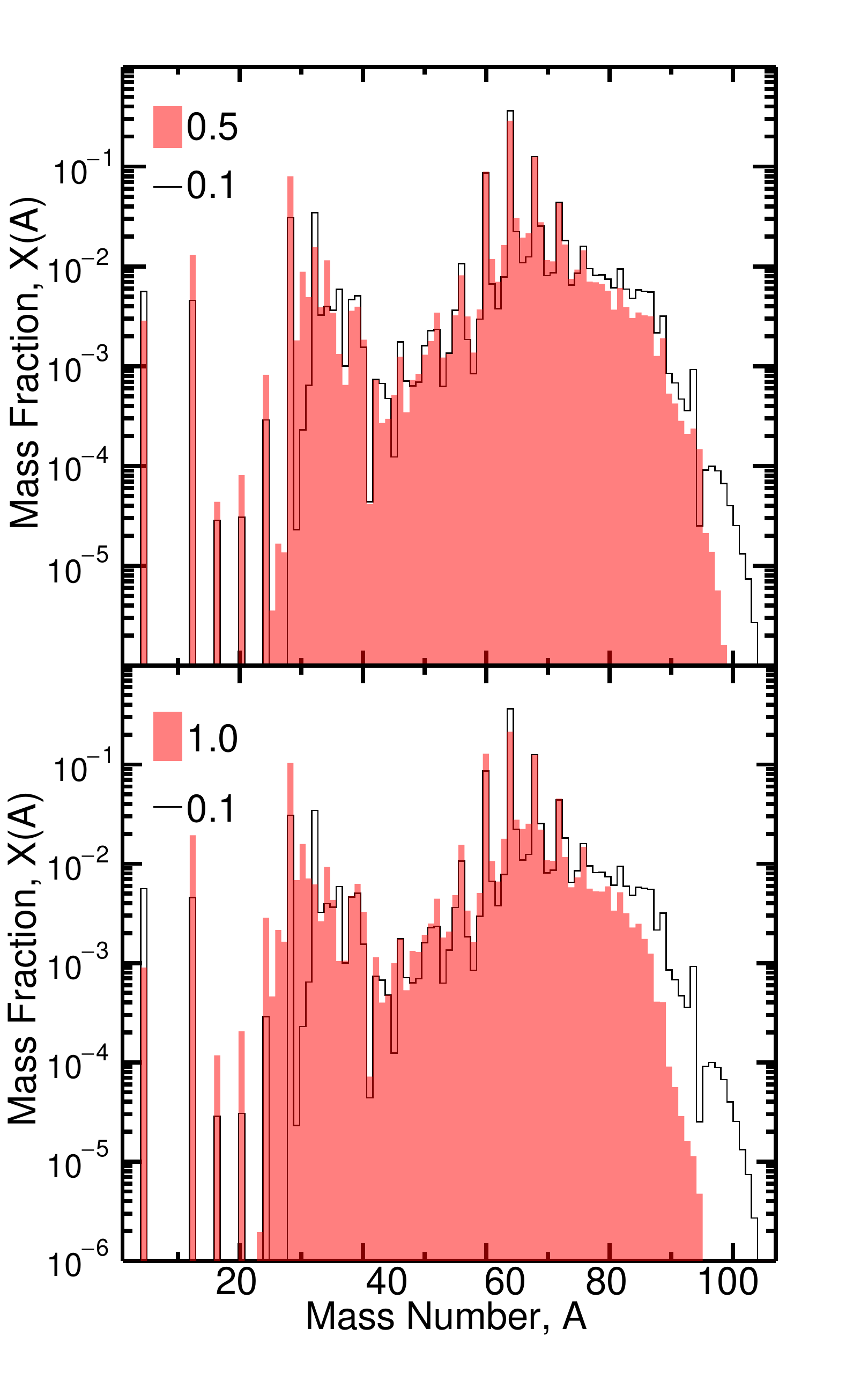}}
 \caption{Ash compositions for model calculations corresponding to the
 light curves in Figure~\ref{fig:LCastro}.}
 \label{fig:AshAstro}
\end{figure*}

\begin{figure}[ht]
\begin{center}
\includegraphics[width=0.75\columnwidth,angle=0]{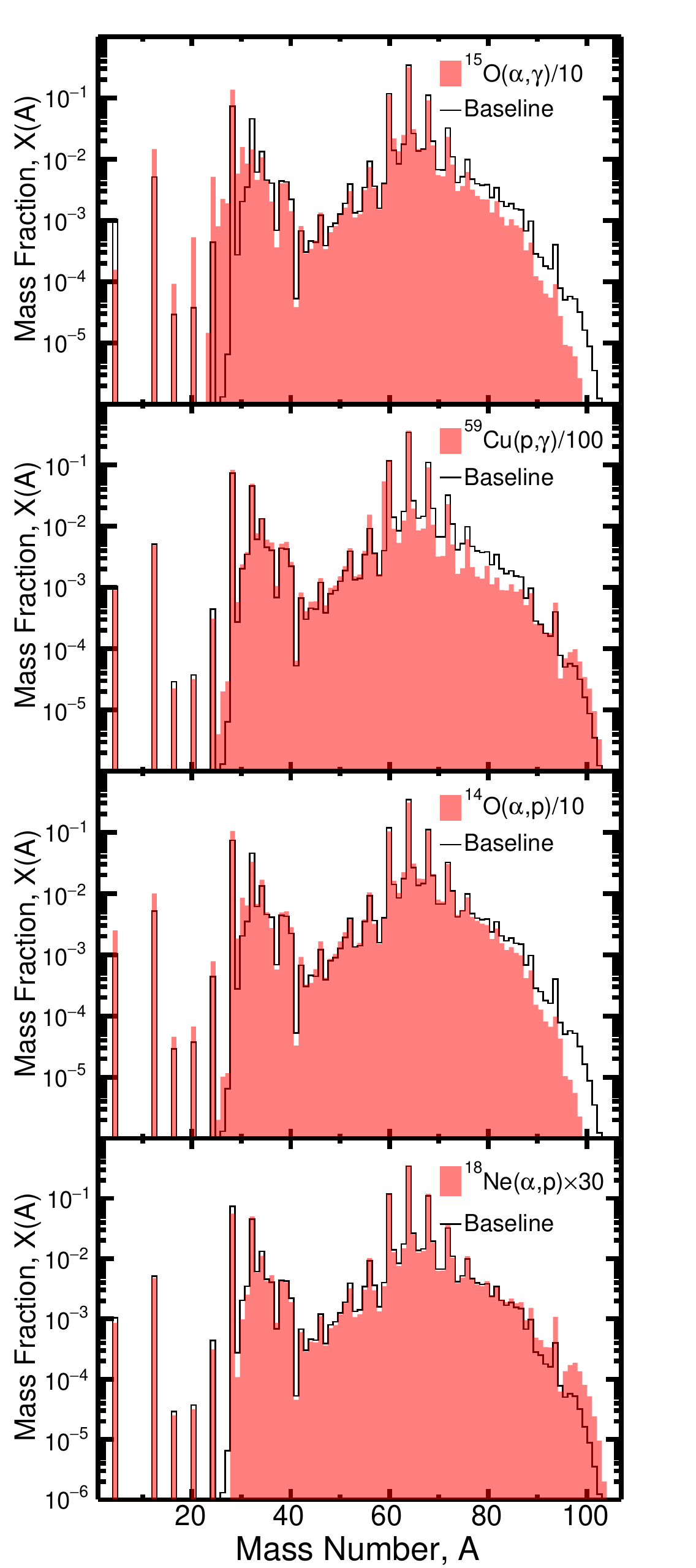}
\caption{Ash compositions for example nuclear reaction rate
variations.
\label{fig:AshRxn}}
\end{center}
\end{figure}

\begin{figure}[ht!]
  \centering
  \subfigure[Baseline.]{\label{fig:flowBase}\includegraphics[width=1\columnwidth]{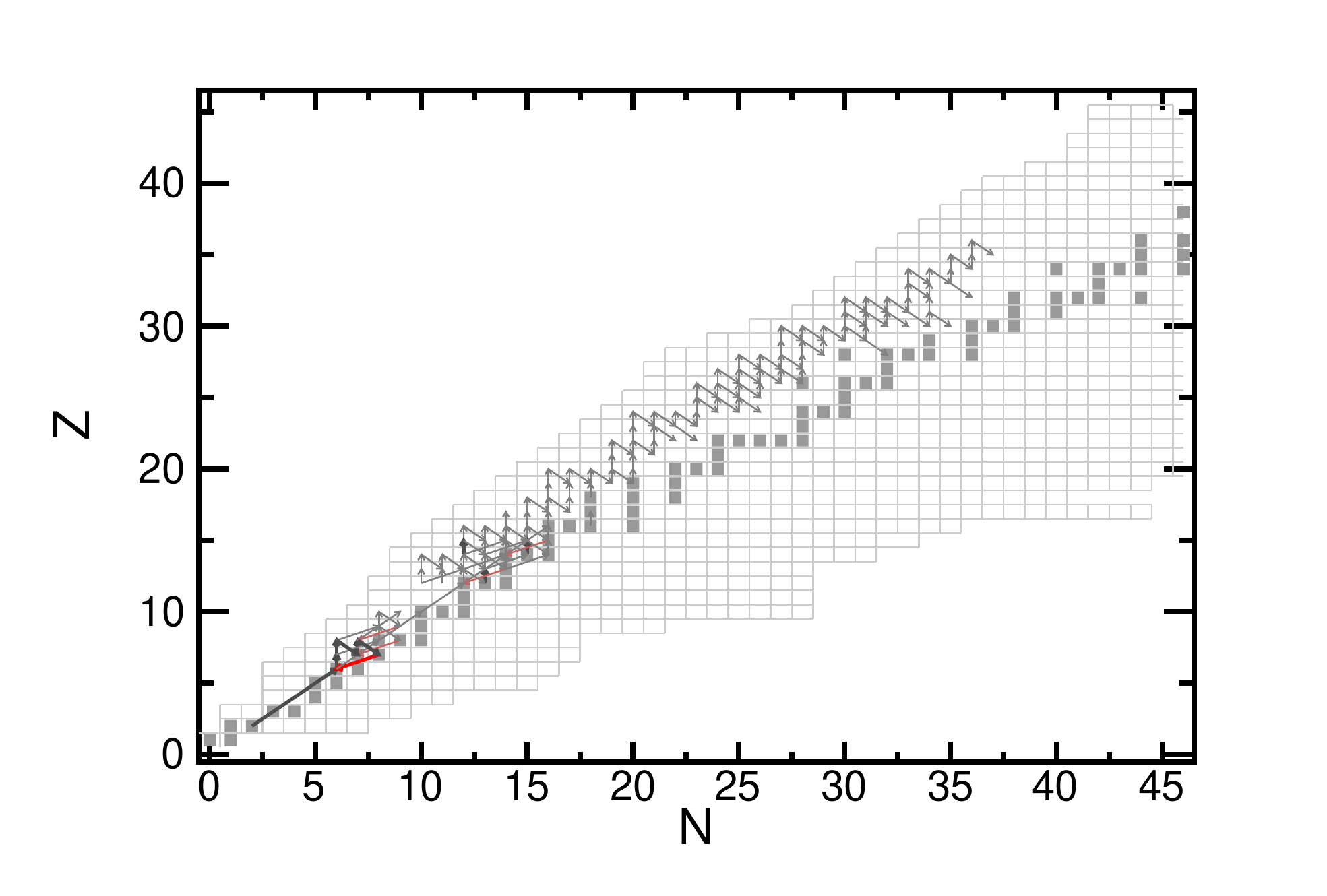}}
  \qquad
  \subfigure[
  $X(H)=0.75$.]{\label{fig:flowXH75}\includegraphics[width=1\columnwidth]{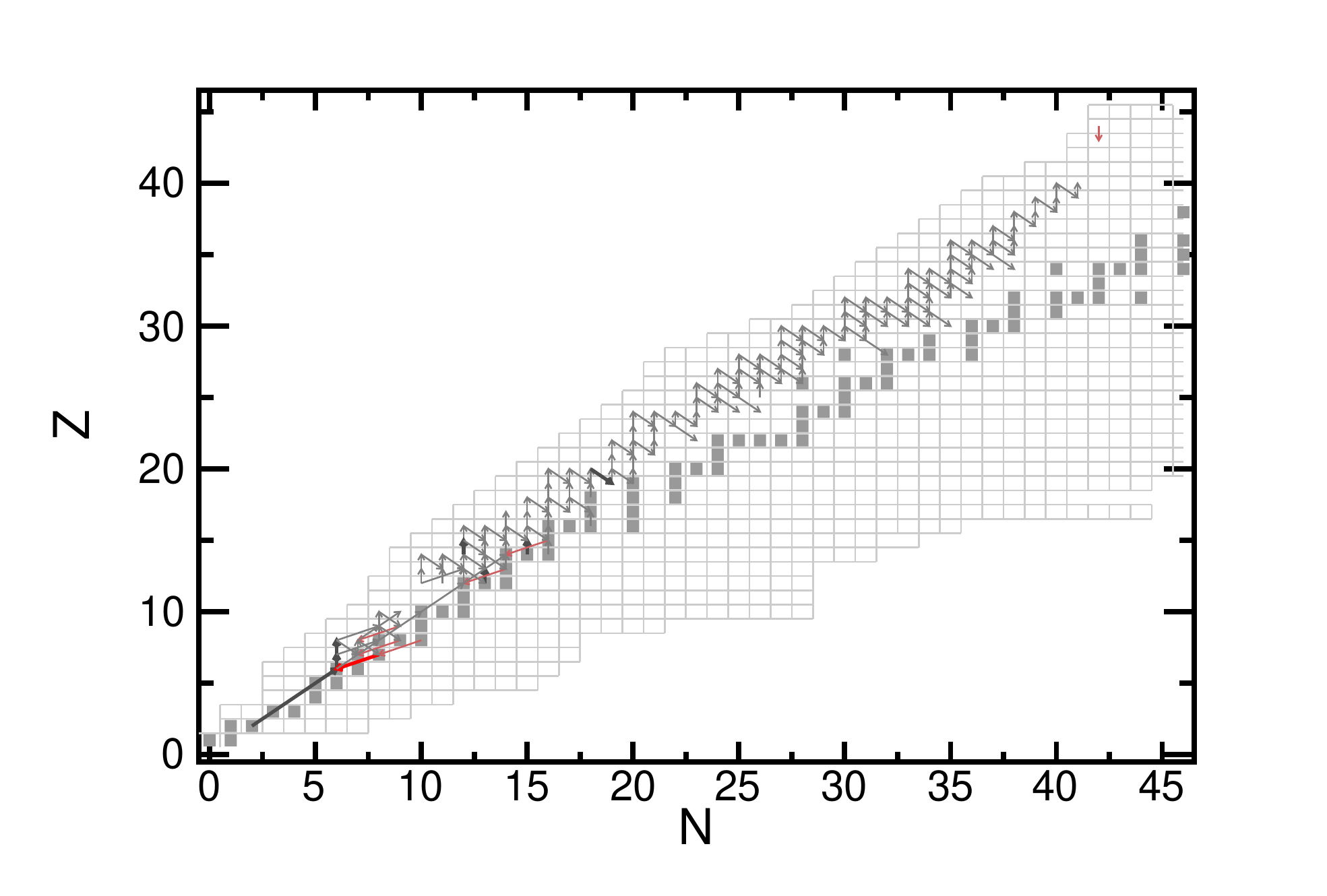}}
  \qquad
  \subfigure[$\dot{M}=0.07~\dot{M}_{\rm
  E}$.]{\label{fig:flow07mdot}\includegraphics[width=1\columnwidth]{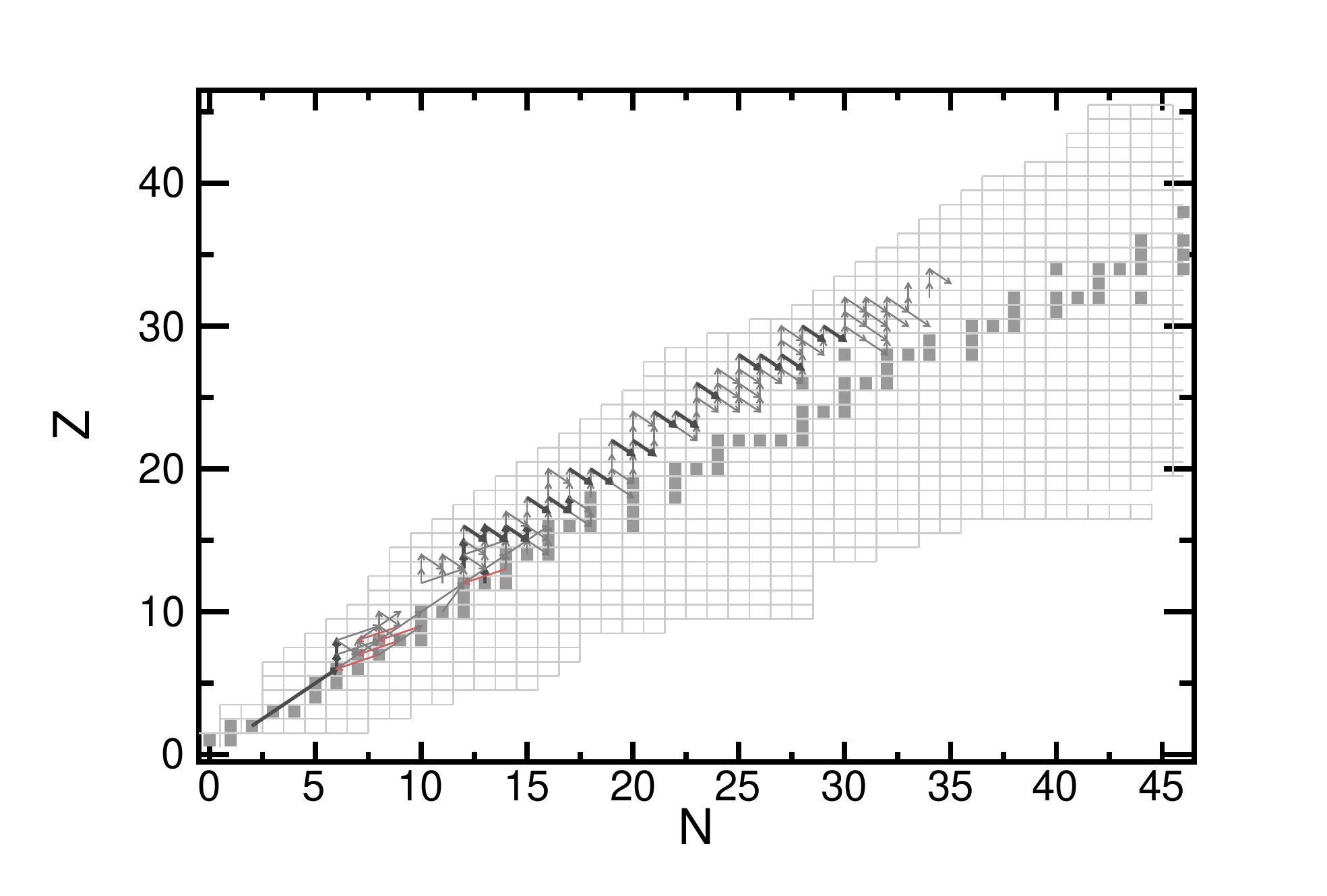}}
 \caption{$\mathcal{F}_{ij}$ for example cases, where the arrow
 indicates the direction $i\rightarrow j$. Thick lines
 correspond to $\mathcal{F}_{ij}>1$~mol/g, whereas thin lines correspond
 to $10^{-1.5}<\mathcal{F}_{ij}<1$~mol/g. $(p,\alpha)$, $(\gamma,p)$,
 and $(\gamma,\alpha)$ reactions are indicated with red lines for
 convenience.}
 \label{fig:flow}
\end{figure}

\begin{table*}[h]
\caption{\label{tab:AshAstro}Ash compositions displayed in Figure~\ref{fig:AshAstro}.
$A$ which do not exceed $X(A)=10^{-10}$ for any conditions are not
listed.}
\resizebox{2.\columnwidth}{!}{\begin{tabular}{c|ccccc|ccc|cccccc}
 & \multicolumn{14}{c}{$X(A)$}\\
 \hline
 & \multicolumn{5}{c}{Vary $\dot{M}$ (in $M_{\rm E}$)} &
 \multicolumn{3}{|c}{Vary $Q_{\rm b}$ (in MeV/u)} &
 \multicolumn{6}{|c}{Vary $X(H)$}\\
 \hline
A	&	0.05	&	0.07	&	0.11	&	0.15	&	0.17	&	0.1	&	0.5	&	1.0	&	0.50	&	0.55	&	0.60	&	0.65	&	0.70	&	0.75	\\
\hline
4	&	3.37E-04	&	1.53E-04	&	2.07E-04	&	9.86E-04	&	1.04E-03	&	5.60E-03	&	2.86E-03	&	9.01E-04	&	1.06E-03	&	7.89E-04	&	6.63E-04	&	9.86E-04	&	1.05E-03	&	1.34E-03	\\
12	&	1.60E-02	&	2.16E-02	&	8.72E-03	&	4.98E-03	&	5.12E-03	&	4.57E-03	&	1.31E-02	&	1.94E-02	&	1.33E-02	&	1.18E-02	&	8.98E-03	&	5.32E-03	&	5.18E-03	&	4.65E-03	\\
13	&	1.97E-10	&	4.55E-10	&	6.89E-12	&	4.71E-14	&	3.10E-14	&	8.06E-15	&	2.34E-11	&	4.10E-12	&	3.27E-10	&	2.34E-10	&	2.58E-10	&	3.13E-12	&	1.85E-14	&	3.10E-14	\\
16	&	5.77E-05	&	9.89E-05	&	3.70E-05	&	2.89E-05	&	2.90E-05	&	2.86E-05	&	4.39E-05	&	1.18E-04	&	8.46E-05	&	5.01E-05	&	3.39E-05	&	2.75E-05	&	3.17E-05	&	2.26E-05	\\
18	&	6.19E-16	&	2.36E-14	&	3.23E-14	&	1.18E-10	&	2.13E-20	&	2.91E-21	&	2.76E-19	&	2.96E-15	&	1.72E-15	&	7.20E-15	&	1.50E-16	&	2.44E-16	&	1.97E-20	&	1.78E-18	\\
19	&	2.66E-14	&	1.33E-11	&	9.24E-12	&	4.51E-10	&	1.02E-20	&	3.03E-24	&	4.78E-20	&	6.86E-16	&	4.23E-18	&	3.41E-13	&	5.35E-16	&	2.90E-19	&	4.39E-22	&	1.71E-20	\\
20	&	1.29E-04	&	1.25E-03	&	1.12E-04	&	3.71E-05	&	3.73E-05	&	3.05E-05	&	8.09E-05	&	2.06E-04	&	9.89E-05	&	7.91E-05	&	6.00E-05	&	3.80E-05	&	4.06E-05	&	2.90E-05	\\
21	&	2.51E-09	&	7.61E-09	&	1.71E-09	&	1.35E-09	&	1.21E-12	&	5.99E-13	&	1.50E-11	&	2.92E-10	&	1.78E-09	&	1.89E-09	&	7.95E-10	&	1.21E-10	&	8.86E-13	&	3.54E-12	\\
22	&	4.07E-13	&	1.66E-07	&	9.29E-08	&	7.62E-09	&	4.38E-15	&	2.12E-17	&	4.69E-15	&	1.32E-11	&	3.04E-14	&	5.73E-11	&	1.32E-13	&	2.30E-15	&	1.83E-16	&	4.78E-15	\\
23	&	8.07E-08	&	1.23E-06	&	2.21E-07	&	1.33E-08	&	1.65E-10	&	5.46E-11	&	3.34E-09	&	1.95E-06	&	1.57E-08	&	2.49E-08	&	1.59E-08	&	3.09E-09	&	7.40E-11	&	2.48E-09	\\
24	&	7.14E-04	&	5.07E-03	&	7.58E-04	&	3.93E-04	&	4.41E-04	&	2.89E-04	&	8.20E-04	&	2.86E-03	&	6.81E-04	&	4.88E-04	&	4.37E-04	&	3.61E-04	&	4.75E-04	&	2.57E-04	\\
25	&	3.59E-06	&	7.17E-05	&	3.10E-05	&	5.24E-07	&	4.35E-07	&	1.23E-07	&	3.57E-06	&	4.62E-04	&	6.53E-07	&	6.70E-07	&	1.73E-06	&	5.46E-07	&	4.45E-07	&	1.85E-06	\\
26	&	7.73E-05	&	4.33E-04	&	2.30E-04	&	3.23E-06	&	1.31E-06	&	2.14E-10	&	1.67E-05	&	2.15E-03	&	4.14E-07	&	1.52E-05	&	1.96E-05	&	6.35E-06	&	2.37E-07	&	1.96E-05	\\
27	&	8.75E-05	&	4.43E-04	&	1.94E-04	&	6.81E-06	&	6.45E-06	&	4.20E-07	&	1.36E-05	&	1.64E-03	&	3.92E-06	&	1.34E-05	&	2.11E-05	&	6.92E-06	&	1.68E-06	&	2.36E-05	\\
28	&	1.78E-01	&	1.64E-01	&	1.14E-01	&	7.92E-02	&	7.34E-02	&	3.08E-02	&	8.00E-02	&	1.04E-01	&	1.47E-01	&	1.38E-01	&	1.27E-01	&	9.34E-02	&	7.36E-02	&	4.66E-02	\\
29	&	2.00E-03	&	2.88E-03	&	1.21E-03	&	3.60E-04	&	2.75E-04	&	2.30E-05	&	1.81E-03	&	6.85E-03	&	5.38E-04	&	9.31E-04	&	1.08E-03	&	5.63E-04	&	2.67E-04	&	3.50E-04	\\
30	&	6.91E-03	&	7.14E-03	&	3.91E-03	&	2.03E-03	&	2.03E-03	&	2.30E-04	&	8.84E-03	&	1.58E-02	&	5.71E-03	&	5.28E-03	&	4.60E-03	&	2.59E-03	&	2.06E-03	&	1.15E-03	\\
31	&	6.80E-03	&	4.98E-03	&	4.69E-03	&	3.65E-03	&	3.48E-03	&	6.44E-04	&	4.93E-03	&	7.11E-03	&	8.61E-03	&	6.89E-03	&	5.69E-03	&	4.42E-03	&	3.51E-03	&	1.76E-03	\\
32	&	7.02E-02	&	3.42E-02	&	4.56E-02	&	4.77E-02	&	4.54E-02	&	3.45E-02	&	1.55E-02	&	6.18E-03	&	1.01E-01	&	7.83E-02	&	6.15E-02	&	5.63E-02	&	4.53E-02	&	2.97E-02	\\
33	&	8.39E-03	&	5.51E-03	&	6.94E-03	&	6.52E-03	&	6.13E-03	&	3.24E-03	&	3.91E-03	&	2.64E-03	&	1.33E-02	&	1.06E-02	&	8.87E-03	&	8.23E-03	&	6.18E-03	&	3.61E-03	\\
34	&	2.54E-02	&	1.18E-02	&	1.30E-02	&	1.33E-02	&	1.33E-02	&	3.95E-03	&	1.15E-02	&	9.31E-03	&	3.61E-02	&	2.50E-02	&	1.77E-02	&	1.54E-02	&	1.34E-02	&	6.93E-03	\\
35	&	1.11E-02	&	6.74E-03	&	5.29E-03	&	4.77E-03	&	4.55E-03	&	3.64E-03	&	3.43E-03	&	4.33E-03	&	1.47E-02	&	1.05E-02	&	7.90E-03	&	6.64E-03	&	4.62E-03	&	3.30E-03	\\
36	&	6.72E-03	&	3.81E-03	&	3.95E-03	&	4.20E-03	&	4.07E-03	&	5.88E-03	&	1.32E-03	&	1.05E-03	&	1.03E-02	&	7.48E-03	&	5.87E-03	&	5.65E-03	&	4.08E-03	&	3.89E-03	\\
37	&	1.46E-03	&	9.90E-04	&	7.23E-04	&	7.40E-04	&	6.89E-04	&	1.00E-03	&	6.50E-04	&	1.06E-03	&	2.13E-03	&	1.64E-03	&	1.21E-03	&	9.89E-04	&	6.89E-04	&	5.52E-04	\\
38	&	1.20E-02	&	6.47E-03	&	4.41E-03	&	4.27E-03	&	4.37E-03	&	4.64E-03	&	3.60E-03	&	4.70E-03	&	1.50E-02	&	1.04E-02	&	7.00E-03	&	5.63E-03	&	4.45E-03	&	3.39E-03	\\
39	&	1.40E-02	&	9.24E-03	&	4.86E-03	&	4.09E-03	&	4.23E-03	&	5.06E-03	&	3.95E-03	&	6.26E-03	&	1.78E-02	&	1.33E-02	&	9.09E-03	&	6.26E-03	&	4.30E-03	&	3.76E-03	\\
40	&	4.68E-03	&	3.49E-03	&	2.61E-03	&	2.25E-03	&	2.21E-03	&	1.54E-03	&	1.84E-03	&	3.28E-03	&	6.54E-03	&	4.77E-03	&	3.61E-03	&	3.12E-03	&	2.23E-03	&	1.44E-03	\\
41	&	1.09E-04	&	8.32E-05	&	7.07E-05	&	6.09E-05	&	5.28E-05	&	4.38E-05	&	4.16E-05	&	7.17E-05	&	1.36E-04	&	1.07E-04	&	8.89E-05	&	8.13E-05	&	5.24E-05	&	2.66E-05	\\
42	&	2.31E-03	&	1.37E-03	&	7.51E-04	&	6.41E-04	&	6.73E-04	&	7.36E-04	&	7.28E-04	&	1.15E-03	&	2.70E-03	&	1.98E-03	&	1.32E-03	&	9.08E-04	&	6.89E-04	&	5.12E-04	\\
43	&	8.42E-04	&	5.96E-04	&	3.73E-04	&	3.11E-04	&	3.03E-04	&	6.71E-04	&	2.70E-04	&	4.00E-04	&	1.04E-03	&	8.29E-04	&	6.03E-04	&	4.60E-04	&	3.05E-04	&	2.60E-04	\\
44	&	1.04E-03	&	7.97E-04	&	5.72E-04	&	5.04E-04	&	4.58E-04	&	4.75E-04	&	2.97E-04	&	4.79E-04	&	1.29E-03	&	1.03E-03	&	7.68E-04	&	7.36E-04	&	4.59E-04	&	3.15E-04	\\
45	&	1.54E-03	&	9.92E-04	&	5.51E-04	&	4.64E-04	&	4.42E-04	&	1.23E-04	&	5.13E-04	&	9.96E-04	&	1.91E-03	&	1.43E-03	&	9.81E-04	&	6.62E-04	&	4.47E-04	&	2.59E-04	\\
46	&	3.92E-03	&	2.41E-03	&	1.32E-03	&	1.13E-03	&	1.21E-03	&	1.75E-03	&	1.24E-03	&	1.74E-03	&	4.85E-03	&	3.55E-03	&	2.29E-03	&	1.58E-03	&	1.24E-03	&	1.03E-03	\\
47	&	1.12E-03	&	8.00E-04	&	4.86E-04	&	4.09E-04	&	3.88E-04	&	7.10E-04	&	3.46E-04	&	5.32E-04	&	1.49E-03	&	1.14E-03	&	7.93E-04	&	6.08E-04	&	3.90E-04	&	3.22E-04	\\
48	&	2.48E-03	&	1.60E-03	&	9.85E-04	&	8.53E-04	&	7.93E-04	&	6.34E-04	&	7.29E-04	&	1.32E-03	&	3.46E-03	&	2.40E-03	&	1.58E-03	&	1.18E-03	&	7.98E-04	&	4.98E-04	\\
49	&	3.02E-03	&	1.81E-03	&	1.09E-03	&	9.40E-04	&	8.96E-04	&	6.93E-04	&	8.38E-04	&	1.29E-03	&	3.79E-03	&	2.73E-03	&	1.76E-03	&	1.29E-03	&	9.06E-04	&	5.84E-04	\\
50	&	4.07E-03	&	2.64E-03	&	1.45E-03	&	1.25E-03	&	1.26E-03	&	1.60E-03	&	1.30E-03	&	1.92E-03	&	5.26E-03	&	3.94E-03	&	2.55E-03	&	1.74E-03	&	1.29E-03	&	1.03E-03	\\
51	&	6.01E-03	&	4.18E-03	&	2.17E-03	&	1.86E-03	&	1.90E-03	&	2.27E-03	&	1.78E-03	&	2.50E-03	&	7.62E-03	&	6.12E-03	&	4.00E-03	&	2.63E-03	&	1.93E-03	&	1.64E-03	\\
52	&	9.15E-03	&	7.06E-03	&	5.01E-03	&	4.20E-03	&	3.91E-03	&	2.34E-03	&	3.44E-03	&	4.44E-03	&	9.50E-03	&	8.91E-03	&	7.12E-03	&	5.34E-03	&	3.92E-03	&	2.13E-03	\\
53	&	3.25E-03	&	2.53E-03	&	1.76E-03	&	1.43E-03	&	1.33E-03	&	6.29E-04	&	1.22E-03	&	1.80E-03	&	3.44E-03	&	3.17E-03	&	2.55E-03	&	1.86E-03	&	1.34E-03	&	6.95E-04	\\
54	&	3.37E-03	&	2.59E-03	&	1.69E-03	&	1.42E-03	&	1.40E-03	&	1.35E-03	&	1.32E-03	&	2.08E-03	&	4.03E-03	&	3.40E-03	&	2.58E-03	&	1.90E-03	&	1.42E-03	&	9.37E-04	\\
55	&	8.27E-03	&	6.40E-03	&	4.18E-03	&	3.53E-03	&	3.46E-03	&	3.63E-03	&	3.24E-03	&	4.83E-03	&	1.01E-02	&	8.36E-03	&	6.34E-03	&	4.68E-03	&	3.50E-03	&	2.39E-03	\\
56	&	3.28E-02	&	2.55E-02	&	1.16E-02	&	9.27E-03	&	9.17E-03	&	1.07E-02	&	8.14E-03	&	1.55E-02	&	4.25E-02	&	3.46E-02	&	2.45E-02	&	1.51E-02	&	9.27E-03	&	9.00E-03	\\
57	&	6.17E-03	&	5.23E-03	&	4.49E-03	&	3.83E-03	&	3.61E-03	&	1.85E-03	&	3.15E-03	&	3.38E-03	&	5.19E-03	&	5.45E-03	&	5.31E-03	&	4.33E-03	&	3.62E-03	&	1.74E-03	\\
58	&	2.00E-03	&	1.89E-03	&	1.81E-03	&	1.62E-03	&	1.58E-03	&	8.46E-04	&	1.37E-03	&	1.64E-03	&	1.87E-03	&	1.90E-03	&	1.96E-03	&	1.73E-03	&	1.59E-03	&	7.92E-04	\\
59	&	7.51E-03	&	6.13E-03	&	4.92E-03	&	4.20E-03	&	4.00E-03	&	2.96E-03	&	3.70E-03	&	5.07E-03	&	8.21E-03	&	7.31E-03	&	6.37E-03	&	5.09E-03	&	4.02E-03	&	2.14E-03	\\
60	&	3.45E-01	&	2.82E-01	&	1.72E-01	&	1.34E-01	&	1.17E-01	&	8.60E-02	&	8.91E-02	&	1.28E-01	&	3.36E-01	&	3.23E-01	&	2.71E-01	&	1.97E-01	&	1.17E-01	&	7.12E-02	\\
61	&	1.42E-02	&	1.74E-02	&	1.77E-02	&	1.49E-02	&	1.40E-02	&	6.67E-03	&	1.19E-02	&	1.07E-02	&	1.08E-02	&	1.34E-02	&	1.56E-02	&	1.52E-02	&	1.40E-02	&	5.91E-03	\\
62	&	3.22E-03	&	6.84E-03	&	9.56E-03	&	8.54E-03	&	8.42E-03	&	3.78E-03	&	7.02E-03	&	6.62E-03	&	1.89E-03	&	3.24E-03	&	5.69E-03	&	7.24E-03	&	8.45E-03	&	3.52E-03	\\
63	&	4.70E-03	&	8.91E-03	&	1.66E-02	&	1.66E-02	&	1.75E-02	&	7.82E-03	&	1.63E-02	&	1.79E-02	&	4.99E-03	&	5.44E-03	&	1.03E-02	&	1.32E-02	&	1.75E-02	&	7.51E-03	\\
64	&	1.50E-01	&	2.69E-01	&	3.44E-01	&	3.52E-01	&	3.41E-01	&	3.63E-01	&	2.87E-01	&	2.14E-01	&	1.19E-01	&	1.95E-01	&	2.81E-01	&	3.52E-01	&	3.39E-01	&	2.50E-01	\\
65	&	4.80E-03	&	8.48E-03	&	1.97E-02	&	2.37E-02	&	2.62E-02	&	2.23E-02	&	3.07E-02	&	2.78E-02	&	3.35E-03	&	5.80E-03	&	9.89E-03	&	1.61E-02	&	2.64E-02	&	2.29E-02	\\
66	&	8.72E-04	&	2.37E-03	&	8.23E-03	&	1.11E-02	&	1.36E-02	&	1.09E-02	&	1.94E-02	&	2.24E-02	&	5.91E-04	&	1.07E-03	&	2.78E-03	&	5.82E-03	&	1.37E-02	&	1.51E-02	\\
67	&	9.91E-04	&	2.59E-03	&	8.93E-03	&	1.25E-02	&	1.46E-02	&	1.24E-02	&	2.14E-02	&	2.53E-02	&	7.17E-04	&	1.38E-03	&	3.32E-03	&	7.00E-03	&	1.47E-02	&	1.51E-02	\\
68	&	1.03E-02	&	2.95E-02	&	8.01E-02	&	1.05E-01	&	1.11E-01	&	1.26E-01	&	1.26E-01	&	1.25E-01	&	8.77E-03	&	2.07E-02	&	4.03E-02	&	7.62E-02	&	1.11E-01	&	1.24E-01	\\
69	&	4.50E-04	&	1.58E-03	&	9.92E-03	&	1.62E-02	&	1.95E-02	&	2.54E-02	&	2.77E-02	&	2.20E-02	&	2.09E-04	&	7.59E-04	&	2.20E-03	&	6.87E-03	&	1.97E-02	&	2.62E-02	\\
70	&	8.40E-05	&	3.39E-04	&	2.77E-03	&	5.06E-03	&	6.72E-03	&	8.08E-03	&	1.15E-02	&	1.08E-02	&	4.19E-05	&	1.55E-04	&	5.41E-04	&	1.78E-03	&	6.81E-03	&	1.24E-02	\\
71	&	9.90E-05	&	3.93E-04	&	2.91E-03	&	5.32E-03	&	6.70E-03	&	8.62E-03	&	1.12E-02	&	1.07E-02	&	5.63E-05	&	2.07E-04	&	6.08E-04	&	2.02E-03	&	6.79E-03	&	1.16E-02	\\
72	&	8.46E-04	&	3.62E-03	&	1.84E-02	&	2.83E-02	&	3.22E-02	&	4.38E-02	&	4.31E-02	&	4.54E-02	&	6.48E-04	&	2.72E-03	&	6.81E-03	&	1.61E-02	&	3.23E-02	&	5.94E-02	\\
73	&	7.28E-05	&	3.76E-04	&	4.57E-03	&	8.71E-03	&	1.09E-02	&	1.82E-02	&	1.66E-02	&	1.17E-02	&	2.99E-05	&	2.08E-04	&	7.43E-04	&	2.90E-03	&	1.11E-02	&	2.09E-02	\\
74	&	1.63E-05	&	9.46E-05	&	1.44E-03	&	3.07E-03	&	4.15E-03	&	6.52E-03	&	7.49E-03	&	5.80E-03	&	6.89E-06	&	5.20E-05	&	2.05E-04	&	8.51E-04	&	4.21E-03	&	1.09E-02	\\
75	&	2.06E-05	&	1.17E-04	&	1.83E-03	&	3.92E-03	&	5.19E-03	&	8.53E-03	&	9.30E-03	&	7.29E-03	&	9.63E-06	&	6.91E-05	&	2.57E-04	&	1.11E-03	&	5.26E-03	&	1.34E-02	\\
76	&	8.70E-05	&	5.25E-04	&	4.57E-03	&	8.11E-03	&	9.85E-03	&	1.59E-02	&	1.45E-02	&	1.48E-02	&	5.86E-05	&	4.65E-04	&	1.39E-03	&	3.77E-03	&	9.82E-03	&	2.83E-02	\\
77	&	1.50E-05	&	1.02E-04	&	1.83E-03	&	3.76E-03	&	4.69E-03	&	9.49E-03	&	7.03E-03	&	5.62E-03	&	6.86E-06	&	7.70E-05	&	2.94E-04	&	1.18E-03	&	4.71E-03	&	1.17E-02	\\
78	&	5.89E-06	&	4.77E-05	&	1.25E-03	&	2.95E-03	&	3.95E-03	&	8.15E-03	&	6.93E-03	&	5.29E-03	&	2.42E-06	&	3.43E-05	&	1.53E-04	&	7.33E-04	&	3.97E-03	&	1.25E-02	\\
79	&	5.08E-06	&	4.16E-05	&	1.15E-03	&	2.78E-03	&	3.72E-03	&	8.23E-03	&	6.68E-03	&	5.25E-03	&	2.34E-06	&	3.16E-05	&	1.40E-04	&	6.85E-04	&	3.73E-03	&	1.26E-02	\\
80	&	1.34E-05	&	1.09E-04	&	1.47E-03	&	2.98E-03	&	3.83E-03	&	7.45E-03	&	5.70E-03	&	5.91E-03	&	8.46E-06	&	1.22E-04	&	4.14E-04	&	1.20E-03	&	3.79E-03	&	1.52E-02	\\
81	&	3.25E-06	&	3.06E-05	&	8.12E-04	&	1.85E-03	&	2.36E-03	&	6.10E-03	&	3.69E-03	&	3.38E-03	&	1.58E-06	&	3.00E-05	&	1.31E-04	&	5.55E-04	&	2.35E-03	&	8.26E-03	\\
82	&	2.51E-06	&	2.71E-05	&	1.03E-03	&	2.60E-03	&	3.43E-03	&	9.39E-03	&	6.09E-03	&	5.17E-03	&	1.16E-06	&	2.62E-05	&	1.34E-04	&	6.56E-04	&	3.42E-03	&	1.34E-02	\\
83	&	1.43E-06	&	1.58E-05	&	5.79E-04	&	1.52E-03	&	2.03E-03	&	5.93E-03	&	3.92E-03	&	3.18E-03	&	7.29E-07	&	1.66E-05	&	8.19E-05	&	3.93E-04	&	2.01E-03	&	9.29E-03	\\
84	&	1.91E-06	&	2.16E-05	&	4.70E-04	&	1.23E-03	&	1.68E-03	&	4.80E-03	&	3.04E-03	&	2.28E-03	&	1.17E-06	&	2.98E-05	&	1.16E-04	&	3.81E-04	&	1.66E-03	&	9.09E-03	\\
85	&	1.44E-06	&	1.81E-05	&	5.64E-04	&	1.45E-03	&	1.87E-03	&	5.80E-03	&	3.45E-03	&	2.49E-03	&	7.44E-07	&	2.36E-05	&	1.17E-04	&	4.54E-04	&	1.84E-03	&	9.86E-03	\\
86	&	4.80E-07	&	7.68E-06	&	3.97E-04	&	1.17E-03	&	1.54E-03	&	5.63E-03	&	3.24E-03	&	1.75E-03	&	2.15E-07	&	8.99E-06	&	5.76E-05	&	3.05E-04	&	1.51E-03	&	9.28E-03	\\
87	&	4.55E-07	&	7.59E-06	&	3.72E-04	&	1.13E-03	&	1.49E-03	&	5.54E-03	&	3.17E-03	&	1.24E-03	&	2.08E-07	&	9.84E-06	&	6.22E-05	&	3.11E-04	&	1.45E-03	&	9.83E-03	\\
88	&	3.54E-07	&	5.77E-06	&	1.74E-04	&	5.14E-04	&	6.75E-04	&	2.16E-03	&	1.27E-03	&	4.08E-04	&	1.89E-07	&	9.88E-06	&	4.82E-05	&	1.66E-04	&	6.54E-04	&	5.31E-03	\\
89	&	2.57E-07	&	5.41E-06	&	2.43E-04	&	7.70E-04	&	9.71E-04	&	3.18E-03	&	1.91E-03	&	4.05E-04	&	1.08E-07	&	8.52E-06	&	5.59E-05	&	2.32E-04	&	9.37E-04	&	8.87E-03	\\
90	&	8.12E-08	&	1.78E-06	&	6.91E-05	&	2.27E-04	&	2.82E-04	&	8.50E-04	&	5.32E-04	&	9.06E-05	&	3.79E-08	&	3.21E-06	&	1.84E-05	&	6.91E-05	&	2.72E-04	&	2.99E-03	\\
91	&	5.44E-08	&	1.31E-06	&	6.42E-05	&	2.05E-04	&	2.50E-04	&	6.80E-04	&	4.23E-04	&	5.62E-05	&	2.07E-08	&	2.38E-06	&	1.76E-05	&	6.46E-05	&	2.40E-04	&	2.69E-03	\\
92	&	3.43E-08	&	9.21E-07	&	4.22E-05	&	1.45E-04	&	1.78E-04	&	4.68E-04	&	2.84E-04	&	2.88E-05	&	1.45E-08	&	1.72E-06	&	1.20E-05	&	4.33E-05	&	1.71E-04	&	2.08E-03	\\
93	&	3.41E-08	&	9.82E-07	&	4.00E-05	&	1.32E-04	&	1.61E-04	&	3.60E-04	&	2.10E-04	&	1.62E-05	&	1.44E-08	&	2.02E-06	&	1.40E-05	&	4.28E-05	&	1.55E-04	&	2.12E-03	\\
94	&	4.67E-08	&	1.85E-06	&	9.27E-05	&	3.35E-04	&	4.00E-04	&	9.24E-04	&	2.37E-04	&	1.14E-05	&	1.69E-08	&	3.95E-06	&	3.34E-05	&	1.09E-04	&	3.82E-04	&	6.93E-03	\\
95	&	1.20E-09	&	2.93E-08	&	1.31E-05	&	5.47E-05	&	7.80E-05	&	2.51E-05	&	1.48E-04	&	4.75E-06	&	2.15E-10	&	8.36E-08	&	1.29E-06	&	7.58E-06	&	7.45E-05	&	1.02E-03	\\
96	&	3.88E-09	&	1.91E-07	&	1.03E-05	&	4.09E-05	&	5.04E-05	&	9.13E-05	&	2.12E-05	&	4.97E-07	&	1.28E-09	&	4.27E-07	&	3.77E-06	&	1.20E-05	&	4.81E-05	&	1.01E-03	\\
97	&	3.06E-09	&	1.92E-07	&	1.13E-05	&	4.77E-05	&	5.73E-05	&	9.96E-05	&	1.38E-05	&	2.50E-07	&	8.95E-10	&	4.24E-07	&	4.14E-06	&	1.36E-05	&	5.46E-05	&	1.38E-03	\\
98	&	1.65E-09	&	1.48E-07	&	9.70E-06	&	4.53E-05	&	5.22E-05	&	8.92E-05	&	5.67E-06	&	8.33E-08	&	3.74E-10	&	2.87E-07	&	3.31E-06	&	1.23E-05	&	4.97E-05	&	1.87E-03	\\
99	&	3.36E-10	&	4.80E-08	&	4.89E-06	&	2.80E-05	&	3.18E-05	&	6.71E-05	&	1.60E-06	&	2.25E-08	&	5.81E-11	&	7.25E-08	&	1.15E-06	&	5.98E-06	&	3.02E-05	&	1.61E-03	\\
100	&	8.86E-11	&	1.52E-08	&	2.13E-06	&	1.39E-05	&	1.62E-05	&	4.01E-05	&	3.20E-07	&	4.44E-09	&	1.23E-11	&	2.00E-08	&	3.63E-07	&	2.53E-06	&	1.54E-05	&	1.10E-03	\\
101	&	5.60E-11	&	1.02E-08	&	9.69E-07	&	6.98E-06	&	8.75E-06	&	2.53E-05	&	5.15E-08	&	6.77E-10	&	6.75E-12	&	1.40E-08	&	1.78E-07	&	1.19E-06	&	8.31E-06	&	1.06E-03	\\
102	&	1.09E-11	&	2.82E-09	&	3.04E-07	&	2.50E-06	&	3.51E-06	&	1.32E-05	&	6.67E-09	&	7.93E-11	&	1.00E-12	&	3.68E-09	&	4.09E-08	&	3.59E-07	&	3.33E-06	&	7.39E-04	\\
103	&	1.43E-12	&	5.22E-10	&	7.46E-08	&	7.32E-07	&	1.23E-06	&	7.39E-06	&	8.91E-10	&	7.90E-12	&	9.27E-14	&	6.87E-10	&	6.05E-09	&	8.19E-08	&	1.17E-06	&	4.88E-04	\\
104	&	1.08E-13	&	4.99E-11	&	1.16E-08	&	1.29E-07	&	2.37E-07	&	2.70E-06	&	9.47E-11	&	5.56E-13	&	5.91E-15	&	8.12E-11	&	5.59E-10	&	1.19E-08	&	2.25E-07	&	1.73E-04	\\
105	&	1.52E-15	&	8.33E-13	&	3.01E-10	&	4.82E-09	&	1.12E-08	&	3.87E-07	&	3.87E-12	&	9.43E-15	&	3.09E-17	&	1.73E-12	&	4.81E-12	&	2.66E-10	&	1.06E-08	&	1.93E-05	\\
\hline
$Q_{\rm imp}^{\rm inner}$ & 13.0 & 15.7 & 15.9 & 14.4 & 13.4 & 11.2
& 11.5 & 13.5 & 10.1 & 13.0 & 15.1 & 15.9 & 15.7 & 14.7 \\ 
\end{tabular}}
\end{table*}

\begin{table*}[h]
\caption{\label{tab:AshNucA} Ash compositions corresponding to the rate variation
numbers of Section~\ref{sec:rates}.
$A$ which never exceed $X(A)=10^{-10}$ for any rate variation are not
listed.}
\resizebox{2.\columnwidth}{!}{\begin{tabular}{c|cccccccccccccc}
A	&	1	&	2	&	3	&	4	&	5	&	6	&	7	&	8	&	9	&	10	&	11	&	12	&	13	&	14	\\
\hline																													\\
4	&	1.56E-04	&	2.10E-03	&	9.11E-04	&	9.57E-04	&	9.53E-04	&	2.48E-03	&	8.88E-04	&	8.60E-04	&	1.12E-03	&	9.80E-04	&	1.22E-03	&	8.79E-04	&	1.19E-03	&	9.03E-04	\\
12	&	1.46E-02	&	4.09E-03	&	4.86E-03	&	6.45E-03	&	4.03E-03	&	1.00E-02	&	5.65E-03	&	4.77E-03	&	4.79E-03	&	4.75E-03	&	3.85E-03	&	4.79E-03	&	4.79E-03	&	5.00E-03	\\
13	&	2.11E-10	&	1.34E-12	&	3.08E-12	&	7.17E-14	&	9.04E-15	&	3.90E-12	&	2.48E-13	&	8.44E-15	&	3.66E-15	&	2.67E-13	&	8.59E-14	&	4.06E-13	&	1.12E-13	&	1.29E-11	\\
16	&	9.20E-05	&	1.93E-05	&	2.26E-05	&	3.90E-05	&	2.34E-05	&	4.56E-05	&	3.41E-05	&	2.51E-05	&	2.70E-05	&	2.31E-05	&	2.67E-05	&	2.64E-05	&	2.63E-05	&	2.91E-05	\\
19	&	3.76E-08	&	3.32E-14	&	1.66E-16	&	5.07E-18	&	2.97E-14	&	5.97E-20	&	1.76E-17	&	3.75E-17	&	4.73E-17	&	1.78E-19	&	3.32E-17	&	2.62E-13	&	6.54E-14	&	2.93E-12	\\
20	&	5.31E-04	&	3.82E-05	&	3.20E-05	&	5.18E-05	&	2.97E-05	&	6.70E-05	&	4.52E-05	&	3.18E-05	&	3.55E-05	&	3.28E-05	&	3.52E-05	&	3.80E-05	&	3.46E-05	&	5.79E-05	\\
21	&	4.82E-09	&	1.57E-09	&	5.15E-10	&	1.42E-12	&	1.26E-10	&	6.53E-12	&	1.18E-10	&	4.22E-10	&	6.20E-12	&	2.18E-12	&	1.67E-10	&	8.21E-10	&	7.37E-11	&	3.21E-09	\\
22	&	4.88E-08	&	2.37E-09	&	4.25E-12	&	1.29E-12	&	2.65E-09	&	2.77E-15	&	7.80E-13	&	1.44E-15	&	1.22E-12	&	1.31E-14	&	1.25E-12	&	2.46E-08	&	2.27E-09	&	1.21E-07	\\
23	&	1.46E-05	&	1.56E-08	&	1.14E-08	&	1.13E-10	&	1.22E-09	&	2.81E-09	&	1.52E-09	&	3.64E-09	&	1.22E-09	&	1.67E-10	&	3.36E-09	&	7.34E-08	&	3.50E-08	&	3.15E-07	\\
24	&	5.14E-03	&	2.31E-04	&	3.10E-04	&	6.06E-04	&	3.30E-04	&	7.83E-04	&	5.53E-04	&	3.12E-04	&	4.16E-04	&	3.68E-04	&	3.87E-04	&	3.73E-04	&	3.74E-04	&	8.36E-05	\\
25	&	8.01E-04	&	1.84E-06	&	4.04E-06	&	6.00E-07	&	3.91E-07	&	2.00E-06	&	7.03E-07	&	2.72E-07	&	9.98E-07	&	3.98E-07	&	1.15E-06	&	8.36E-07	&	2.47E-06	&	2.29E-06	\\
26	&	2.24E-03	&	2.60E-05	&	2.01E-05	&	2.23E-07	&	5.12E-07	&	1.02E-05	&	2.91E-06	&	2.76E-09	&	7.07E-06	&	4.71E-07	&	1.02E-05	&	6.44E-06	&	1.08E-05	&	3.17E-05	\\
27	&	1.90E-03	&	5.33E-05	&	2.94E-05	&	1.07E-06	&	1.06E-06	&	1.16E-05	&	3.36E-06	&	6.13E-07	&	9.68E-06	&	1.09E-06	&	1.65E-05	&	1.92E-05	&	1.29E-05	&	5.37E-05	\\
28	&	1.36E-01	&	7.38E-02	&	8.35E-02	&	7.22E-02	&	7.85E-02	&	1.05E-01	&	7.39E-02	&	5.61E-02	&	7.31E-02	&	6.99E-02	&	7.48E-02	&	7.17E-02	&	7.19E-02	&	7.15E-02	\\
29	&	5.81E-03	&	2.89E-04	&	5.79E-04	&	2.89E-04	&	3.72E-04	&	1.83E-03	&	3.78E-04	&	1.08E-04	&	3.37E-04	&	2.45E-04	&	3.55E-04	&	3.47E-04	&	2.93E-04	&	2.53E-04	\\
30	&	1.57E-02	&	1.62E-03	&	2.40E-03	&	2.20E-03	&	1.82E-03	&	8.54E-03	&	2.45E-03	&	9.87E-04	&	1.85E-03	&	1.77E-03	&	1.96E-03	&	1.80E-03	&	1.65E-03	&	1.61E-03	\\
31	&	8.44E-03	&	3.20E-03	&	3.78E-03	&	3.43E-03	&	3.54E-03	&	6.35E-03	&	3.66E-03	&	2.53E-03	&	3.41E-03	&	3.35E-03	&	3.40E-03	&	3.13E-03	&	3.23E-03	&	3.19E-03	\\
32	&	1.44E-02	&	5.67E-02	&	4.97E-02	&	4.03E-02	&	5.11E-02	&	3.29E-02	&	3.92E-02	&	5.00E-02	&	4.52E-02	&	4.75E-02	&	4.53E-02	&	4.46E-02	&	4.49E-02	&	4.70E-02	\\
33	&	4.59E-03	&	7.22E-03	&	7.69E-03	&	5.70E-03	&	6.01E-03	&	6.77E-03	&	5.78E-03	&	5.88E-03	&	6.49E-03	&	6.43E-03	&	6.24E-03	&	5.98E-03	&	6.02E-03	&	6.22E-03	\\
34	&	1.08E-02	&	1.29E-02	&	1.35E-02	&	1.27E-02	&	1.12E-02	&	1.67E-02	&	1.29E-02	&	1.11E-02	&	1.28E-02	&	1.28E-02	&	1.27E-02	&	1.22E-02	&	1.22E-02	&	1.27E-02	\\
35	&	4.46E-03	&	5.72E-03	&	6.05E-03	&	3.87E-03	&	5.12E-03	&	4.94E-03	&	4.31E-03	&	4.27E-03	&	4.75E-03	&	4.65E-03	&	4.54E-03	&	4.13E-03	&	4.48E-03	&	4.41E-03	\\
36	&	2.03E-03	&	6.09E-03	&	5.44E-03	&	3.30E-03	&	5.12E-03	&	2.72E-03	&	3.45E-03	&	5.28E-03	&	4.26E-03	&	4.47E-03	&	4.12E-03	&	3.99E-03	&	4.18E-03	&	4.24E-03	\\
37	&	3.62E-04	&	1.09E-03	&	1.06E-03	&	5.66E-04	&	7.28E-04	&	5.72E-04	&	5.98E-04	&	8.64E-04	&	7.56E-04	&	7.54E-04	&	7.17E-04	&	6.97E-04	&	7.16E-04	&	7.29E-04	\\
38	&	4.02E-03	&	5.37E-03	&	5.21E-03	&	3.70E-03	&	3.97E-03	&	4.90E-03	&	4.02E-03	&	4.28E-03	&	4.27E-03	&	4.22E-03	&	4.19E-03	&	4.12E-03	&	4.12E-03	&	4.20E-03	\\
39	&	4.08E-03	&	5.98E-03	&	5.58E-03	&	3.34E-03	&	4.63E-03	&	5.12E-03	&	3.87E-03	&	4.04E-03	&	4.07E-03	&	4.03E-03	&	4.00E-03	&	3.88E-03	&	3.92E-03	&	3.94E-03	\\
40	&	1.42E-03	&	2.47E-03	&	2.62E-03	&	1.96E-03	&	2.36E-03	&	2.82E-03	&	2.11E-03	&	1.90E-03	&	2.26E-03	&	2.20E-03	&	2.27E-03	&	2.26E-03	&	2.21E-03	&	2.20E-03	\\
41	&	3.84E-05	&	5.81E-05	&	6.39E-05	&	4.08E-05	&	5.84E-05	&	3.28E-05	&	4.41E-05	&	4.55E-05	&	5.48E-05	&	5.14E-05	&	5.49E-05	&	5.17E-05	&	5.50E-05	&	5.50E-05	\\
42	&	8.05E-04	&	8.79E-04	&	8.28E-04	&	5.66E-04	&	6.30E-04	&	9.16E-04	&	6.29E-04	&	6.05E-04	&	6.16E-04	&	6.11E-04	&	6.14E-04	&	6.00E-04	&	5.97E-04	&	6.08E-04	\\
43	&	2.85E-04	&	4.31E-04	&	4.13E-04	&	2.47E-04	&	3.76E-04	&	3.26E-04	&	2.78E-04	&	2.86E-04	&	3.06E-04	&	3.02E-04	&	3.03E-04	&	2.72E-04	&	3.05E-04	&	2.94E-04	\\
44	&	3.42E-04	&	5.61E-04	&	5.86E-04	&	3.63E-04	&	4.88E-04	&	3.47E-04	&	4.25E-04	&	4.20E-04	&	4.77E-04	&	4.70E-04	&	4.87E-04	&	4.64E-04	&	4.94E-04	&	4.56E-04	\\
45	&	4.69E-04	&	5.95E-04	&	5.98E-04	&	3.57E-04	&	4.81E-04	&	5.78E-04	&	3.87E-04	&	4.08E-04	&	4.26E-04	&	4.16E-04	&	4.25E-04	&	4.14E-04	&	4.14E-04	&	4.21E-04	\\
46	&	1.34E-03	&	1.55E-03	&	1.43E-03	&	1.02E-03	&	1.19E-03	&	1.65E-03	&	1.14E-03	&	1.11E-03	&	1.11E-03	&	1.10E-03	&	1.12E-03	&	1.06E-03	&	1.08E-03	&	1.09E-03	\\
47	&	3.39E-04	&	5.18E-04	&	5.11E-04	&	3.34E-04	&	4.52E-04	&	4.19E-04	&	3.67E-04	&	3.64E-04	&	4.00E-04	&	4.04E-04	&	4.04E-04	&	3.75E-04	&	4.10E-04	&	3.87E-04	\\
48	&	6.41E-04	&	9.81E-04	&	9.90E-04	&	7.13E-04	&	8.18E-04	&	8.58E-04	&	7.53E-04	&	7.04E-04	&	7.98E-04	&	7.86E-04	&	8.02E-04	&	8.01E-04	&	8.05E-04	&	7.78E-04	\\
49	&	8.13E-04	&	1.09E-03	&	1.08E-03	&	8.19E-04	&	9.29E-04	&	1.03E-03	&	8.66E-04	&	7.91E-04	&	8.83E-04	&	8.76E-04	&	8.94E-04	&	8.79E-04	&	8.89E-04	&	8.67E-04	\\
50	&	1.23E-03	&	1.63E-03	&	1.53E-03	&	1.11E-03	&	1.30E-03	&	1.64E-03	&	1.22E-03	&	1.18E-03	&	1.20E-03	&	1.20E-03	&	1.22E-03	&	1.18E-03	&	1.19E-03	&	1.19E-03	\\
51	&	1.62E-03	&	2.51E-03	&	2.24E-03	&	1.71E-03	&	1.91E-03	&	2.32E-03	&	1.86E-03	&	1.83E-03	&	1.79E-03	&	1.80E-03	&	1.84E-03	&	1.82E-03	&	1.80E-03	&	1.80E-03	\\
52	&	2.99E-03	&	4.34E-03	&	4.37E-03	&	3.92E-03	&	3.96E-03	&	3.68E-03	&	4.22E-03	&	3.21E-03	&	3.93E-03	&	3.84E-03	&	3.94E-03	&	4.03E-03	&	3.94E-03	&	3.94E-03	\\
53	&	1.12E-03	&	1.47E-03	&	1.54E-03	&	1.31E-03	&	1.37E-03	&	1.33E-03	&	1.42E-03	&	1.07E-03	&	1.33E-03	&	1.30E-03	&	1.34E-03	&	1.34E-03	&	1.32E-03	&	1.33E-03	\\
54	&	1.26E-03	&	1.63E-03	&	1.62E-03	&	1.31E-03	&	1.42E-03	&	1.56E-03	&	1.45E-03	&	1.20E-03	&	1.37E-03	&	1.35E-03	&	1.39E-03	&	1.36E-03	&	1.36E-03	&	1.37E-03	\\
55	&	3.12E-03	&	4.09E-03	&	3.98E-03	&	3.24E-03	&	3.52E-03	&	3.78E-03	&	3.55E-03	&	3.06E-03	&	3.38E-03	&	3.36E-03	&	3.46E-03	&	3.37E-03	&	3.40E-03	&	3.38E-03	\\
56	&	7.43E-03	&	2.00E-02	&	1.54E-02	&	7.02E-03	&	1.03E-02	&	1.04E-02	&	7.95E-03	&	1.02E-02	&	8.57E-03	&	8.70E-03	&	8.85E-03	&	8.87E-03	&	8.76E-03	&	9.00E-03	\\
57	&	3.41E-03	&	3.63E-03	&	3.65E-03	&	3.87E-03	&	3.56E-03	&	3.18E-03	&	3.98E-03	&	2.98E-03	&	3.58E-03	&	3.56E-03	&	3.64E-03	&	3.59E-03	&	3.60E-03	&	3.61E-03	\\
58	&	1.52E-03	&	1.58E-03	&	1.66E-03	&	1.62E-03	&	1.56E-03	&	1.50E-03	&	1.72E-03	&	1.33E-03	&	1.56E-03	&	1.55E-03	&	1.58E-03	&	1.57E-03	&	1.57E-03	&	1.57E-03	\\
59	&	3.92E-03	&	4.62E-03	&	5.41E-02	&	4.11E-03	&	4.07E-03	&	3.91E-03	&	4.19E-03	&	3.49E-03	&	3.98E-03	&	3.95E-03	&	4.04E-03	&	3.97E-03	&	3.97E-03	&	3.98E-03	\\
60	&	1.12E-01	&	1.58E-01	&	1.18E-01	&	3.42E-01	&	1.32E-01	&	1.03E-01	&	9.75E-02	&	1.20E-01	&	1.17E-01	&	1.15E-01	&	1.18E-01	&	1.16E-01	&	1.18E-01	&	1.20E-01	\\
61	&	2.17E-02	&	1.19E-02	&	9.10E-03	&	3.47E-02	&	1.30E-02	&	1.60E-02	&	1.38E-02	&	1.28E-02	&	1.38E-02	&	1.38E-02	&	1.42E-02	&	1.40E-02	&	1.39E-02	&	1.40E-02	\\
62	&	1.35E-02	&	6.35E-03	&	5.37E-03	&	1.79E-02	&	7.83E-03	&	1.02E-02	&	8.22E-03	&	7.54E-03	&	8.28E-03	&	8.24E-03	&	8.43E-03	&	8.41E-03	&	8.27E-03	&	8.35E-03	\\
63	&	2.49E-02	&	1.29E-02	&	1.23E-02	&	4.00E-02	&	1.63E-02	&	2.22E-02	&	1.78E-02	&	1.49E-02	&	2.98E-02	&	1.70E-02	&	1.72E-02	&	1.71E-02	&	1.67E-02	&	1.70E-02	\\
64	&	3.15E-01	&	3.69E-01	&	3.64E-01	&	1.91E-01	&	3.61E-01	&	3.04E-01	&	3.10E-01	&	3.53E-01	&	3.27E-01	&	3.44E-01	&	3.44E-01	&	3.44E-01	&	3.49E-01	&	3.45E-01	\\
65	&	3.09E-02	&	1.80E-02	&	1.95E-02	&	2.39E-02	&	2.37E-02	&	3.08E-02	&	2.88E-02	&	2.51E-02	&	2.59E-02	&	2.60E-02	&	2.58E-02	&	2.62E-02	&	2.53E-02	&	2.56E-02	\\
66	&	1.46E-02	&	8.19E-03	&	8.55E-03	&	1.49E-02	&	1.14E-02	&	1.75E-02	&	1.69E-02	&	1.27E-02	&	1.37E-02	&	1.35E-02	&	1.32E-02	&	1.35E-02	&	1.26E-02	&	1.29E-02	\\
67	&	1.48E-02	&	8.77E-03	&	9.18E-03	&	1.53E-02	&	1.21E-02	&	1.73E-02	&	1.84E-02	&	1.34E-02	&	1.48E-02	&	1.48E-02	&	1.43E-02	&	1.44E-02	&	1.35E-02	&	1.39E-02	\\
68	&	9.09E-02	&	9.20E-02	&	9.18E-02	&	5.46E-02	&	1.06E-01	&	1.05E-01	&	1.14E-01	&	1.19E-01	&	1.14E-01	&	1.12E-01	&	1.12E-01	&	1.12E-01	&	1.13E-01	&	1.11E-01	\\
69	&	1.68E-02	&	9.87E-03	&	1.06E-02	&	1.26E-02	&	1.46E-02	&	2.08E-02	&	2.62E-02	&	1.84E-02	&	1.99E-02	&	1.98E-02	&	1.92E-02	&	1.99E-02	&	1.87E-02	&	1.91E-02	\\
70	&	5.50E-03	&	3.01E-03	&	3.21E-03	&	5.33E-03	&	4.63E-03	&	7.93E-03	&	1.04E-02	&	6.28E-03	&	6.63E-03	&	6.75E-03	&	6.35E-03	&	6.65E-03	&	6.11E-03	&	6.35E-03	\\
71	&	5.20E-03	&	3.03E-03	&	3.28E-03	&	4.81E-03	&	4.62E-03	&	7.37E-03	&	1.02E-02	&	6.33E-03	&	6.73E-03	&	6.84E-03	&	6.48E-03	&	6.79E-03	&	6.21E-03	&	6.50E-03	\\
72	&	2.31E-02	&	2.37E-02	&	2.31E-02	&	1.44E-02	&	2.84E-02	&	3.02E-02	&	3.67E-02	&	3.58E-02	&	3.32E-02	&	3.25E-02	&	3.21E-02	&	3.28E-02	&	3.26E-02	&	3.22E-02	\\
73	&	8.15E-03	&	4.85E-03	&	5.05E-03	&	5.57E-03	&	7.42E-03	&	1.08E-02	&	1.61E-02	&	1.04E-02	&	1.09E-02	&	1.12E-02	&	1.07E-02	&	1.14E-02	&	1.05E-02	&	1.08E-02	\\
74	&	3.01E-03	&	1.60E-03	&	1.68E-03	&	2.65E-03	&	2.57E-03	&	4.46E-03	&	7.11E-03	&	3.90E-03	&	4.01E-03	&	4.22E-03	&	3.95E-03	&	4.26E-03	&	3.81E-03	&	4.01E-03	\\
75	&	3.65E-03	&	1.98E-03	&	2.07E-03	&	3.08E-03	&	3.17E-03	&	5.39E-03	&	8.88E-03	&	4.87E-03	&	5.05E-03	&	5.31E-03	&	4.99E-03	&	5.42E-03	&	4.79E-03	&	5.08E-03	\\
76	&	6.18E-03	&	6.74E-03	&	6.16E-03	&	4.16E-03	&	8.03E-03	&	8.64E-03	&	1.20E-02	&	1.10E-02	&	9.84E-03	&	9.83E-03	&	9.66E-03	&	1.01E-02	&	9.90E-03	&	9.85E-03	\\
77	&	3.18E-03	&	2.15E-03	&	2.15E-03	&	2.04E-03	&	3.21E-03	&	4.12E-03	&	6.71E-03	&	4.56E-03	&	4.71E-03	&	4.81E-03	&	4.66E-03	&	5.02E-03	&	4.62E-03	&	4.68E-03	\\
78	&	2.51E-03	&	1.49E-03	&	1.51E-03	&	2.09E-03	&	2.38E-03	&	3.57E-03	&	6.48E-03	&	3.72E-03	&	3.88E-03	&	4.02E-03	&	3.85E-03	&	4.23E-03	&	3.76E-03	&	3.92E-03	\\
79	&	2.21E-03	&	1.41E-03	&	1.40E-03	&	1.92E-03	&	2.23E-03	&	3.20E-03	&	6.02E-03	&	3.52E-03	&	3.67E-03	&	3.79E-03	&	3.65E-03	&	4.01E-03	&	3.58E-03	&	3.71E-03	\\
80	&	2.19E-03	&	2.61E-03	&	2.26E-03	&	1.70E-03	&	3.02E-03	&	3.04E-03	&	4.68E-03	&	4.24E-03	&	3.73E-03	&	3.75E-03	&	3.70E-03	&	3.90E-03	&	3.85E-03	&	3.80E-03	\\
81	&	1.35E-03	&	1.18E-03	&	1.13E-03	&	1.08E-03	&	1.66E-03	&	1.81E-03	&	3.20E-03	&	2.42E-03	&	2.40E-03	&	2.40E-03	&	2.35E-03	&	2.55E-03	&	2.42E-03	&	2.37E-03	\\
82	&	2.06E-03	&	1.49E-03	&	1.46E-03	&	1.83E-03	&	2.25E-03	&	2.67E-03	&	4.97E-03	&	3.38E-03	&	3.51E-03	&	3.47E-03	&	3.44E-03	&	3.75E-03	&	3.51E-03	&	3.44E-03	\\
83	&	1.13E-03	&	9.53E-04	&	9.07E-04	&	1.16E-03	&	1.37E-03	&	1.54E-03	&	2.90E-03	&	2.08E-03	&	2.09E-03	&	2.05E-03	&	2.03E-03	&	2.20E-03	&	2.11E-03	&	2.04E-03	\\
84	&	8.22E-04	&	1.06E-03	&	8.97E-04	&	9.49E-04	&	1.24E-03	&	1.19E-03	&	2.15E-03	&	1.90E-03	&	1.69E-03	&	1.65E-03	&	1.65E-03	&	1.76E-03	&	1.74E-03	&	1.68E-03	\\
85	&	1.05E-03	&	1.24E-03	&	1.12E-03	&	1.05E-03	&	1.51E-03	&	1.33E-03	&	2.21E-03	&	2.17E-03	&	1.93E-03	&	1.87E-03	&	1.86E-03	&	1.97E-03	&	2.03E-03	&	1.90E-03	\\
86	&	8.27E-04	&	9.00E-04	&	8.51E-04	&	1.00E-03	&	1.19E-03	&	1.07E-03	&	1.91E-03	&	1.82E-03	&	1.64E-03	&	1.57E-03	&	1.58E-03	&	1.66E-03	&	1.73E-03	&	1.60E-03	\\
87	&	7.52E-04	&	9.99E-04	&	9.36E-04	&	1.01E-03	&	1.24E-03	&	9.80E-04	&	1.68E-03	&	1.89E-03	&	1.56E-03	&	1.51E-03	&	1.51E-03	&	1.54E-03	&	1.68E-03	&	1.54E-03	\\
88	&	3.25E-04	&	5.95E-04	&	5.15E-04	&	4.40E-04	&	6.24E-04	&	4.14E-04	&	6.67E-04	&	9.54E-04	&	6.72E-04	&	6.67E-04	&	6.67E-04	&	6.63E-04	&	7.46E-04	&	6.92E-04	\\
89	&	4.34E-04	&	8.94E-04	&	8.09E-04	&	7.00E-04	&	9.57E-04	&	5.54E-04	&	9.57E-04	&	1.53E-03	&	9.89E-04	&	9.93E-04	&	9.96E-04	&	9.52E-04	&	1.13E-03	&	1.03E-03	\\
90	&	1.22E-04	&	2.87E-04	&	2.54E-04	&	2.07E-04	&	2.92E-04	&	1.52E-04	&	2.72E-04	&	4.90E-04	&	2.89E-04	&	2.92E-04	&	2.96E-04	&	2.75E-04	&	3.33E-04	&	3.07E-04	\\
91	&	1.04E-04	&	2.77E-04	&	2.48E-04	&	1.81E-04	&	2.75E-04	&	1.27E-04	&	2.29E-04	&	4.44E-04	&	2.51E-04	&	2.54E-04	&	2.56E-04	&	2.37E-04	&	2.95E-04	&	2.67E-04	\\
92	&	6.38E-05	&	2.02E-04	&	1.78E-04	&	1.32E-04	&	1.95E-04	&	8.12E-05	&	1.66E-04	&	3.41E-04	&	1.79E-04	&	1.83E-04	&	1.86E-04	&	1.68E-04	&	2.11E-04	&	1.93E-04	\\
93	&	5.47E-05	&	2.13E-04	&	1.80E-04	&	1.14E-04	&	1.88E-04	&	6.64E-05	&	1.41E-04	&	3.37E-04	&	1.58E-04	&	1.62E-04	&	1.65E-04	&	1.48E-04	&	1.89E-04	&	1.73E-04	\\
94	&	9.08E-05	&	6.55E-04	&	5.65E-04	&	2.66E-04	&	5.25E-04	&	9.75E-05	&	3.50E-04	&	1.08E-03	&	3.97E-04	&	4.28E-04	&	4.32E-04	&	3.81E-04	&	4.93E-04	&	4.27E-04	\\
95	&	2.73E-05	&	2.40E-05	&	3.28E-05	&	6.09E-05	&	5.42E-05	&	4.27E-05	&	7.04E-05	&	6.22E-05	&	7.03E-05	&	5.76E-05	&	6.37E-05	&	4.94E-05	&	8.22E-05	&	7.64E-05	\\
96	&	9.25E-06	&	7.97E-05	&	7.01E-05	&	3.34E-05	&	6.29E-05	&	1.04E-05	&	4.40E-05	&	1.36E-04	&	4.88E-05	&	5.15E-05	&	5.24E-05	&	4.51E-05	&	6.14E-05	&	5.25E-05	\\
97	&	8.77E-06	&	9.63E-05	&	8.84E-05	&	3.74E-05	&	7.54E-05	&	9.08E-06	&	4.98E-05	&	1.71E-04	&	5.58E-05	&	6.02E-05	&	6.08E-05	&	5.26E-05	&	7.23E-05	&	6.02E-05	\\
98	&	6.06E-06	&	9.88E-05	&	9.78E-05	&	3.33E-05	&	7.57E-05	&	5.53E-06	&	4.35E-05	&	1.87E-04	&	5.12E-05	&	5.73E-05	&	5.75E-05	&	5.10E-05	&	6.99E-05	&	5.65E-05	\\
99	&	2.65E-06	&	5.65E-05	&	6.19E-05	&	2.15E-05	&	4.79E-05	&	2.28E-06	&	2.76E-05	&	1.35E-04	&	3.27E-05	&	3.84E-05	&	3.78E-05	&	3.29E-05	&	4.64E-05	&	3.60E-05	\\
100	&	9.29E-07	&	2.89E-05	&	3.46E-05	&	1.04E-05	&	2.56E-05	&	7.33E-07	&	1.32E-05	&	8.02E-05	&	1.66E-05	&	2.05E-05	&	1.96E-05	&	1.69E-05	&	2.46E-05	&	1.82E-05	\\
101	&	3.08E-07	&	1.74E-05	&	2.23E-05	&	4.76E-06	&	1.40E-05	&	2.10E-07	&	5.96E-06	&	5.21E-05	&	8.21E-06	&	1.06E-05	&	9.95E-06	&	8.93E-06	&	1.30E-05	&	9.13E-06	\\
102	&	7.51E-08	&	6.58E-06	&	9.54E-06	&	1.61E-06	&	5.38E-06	&	4.65E-08	&	2.04E-06	&	2.42E-05	&	3.04E-06	&	4.04E-06	&	3.69E-06	&	3.44E-06	&	5.07E-06	&	3.30E-06	\\
103	&	1.41E-08	&	1.88E-06	&	3.32E-06	&	4.60E-07	&	1.65E-06	&	8.64E-09	&	6.27E-07	&	9.53E-06	&	9.82E-07	&	1.33E-06	&	1.15E-06	&	1.12E-06	&	1.70E-06	&	9.99E-07	\\
104	&	1.79E-09	&	2.91E-07	&	6.73E-07	&	7.62E-08	&	3.07E-07	&	1.15E-09	&	1.18E-07	&	2.01E-06	&	1.94E-07	&	2.49E-07	&	2.06E-07	&	2.12E-07	&	3.28E-07	&	1.68E-07	\\
105	&	3.27E-11	&	8.23E-09	&	3.07E-08	&	3.05E-09	&	1.05E-08	&	3.39E-11	&	6.63E-09	&	1.01E-07	&	9.62E-09	&	1.16E-08	&	8.45E-09	&	9.43E-09	&	1.58E-08	&	6.52E-09	\\
\hline
$Q_{\rm imp}^{\rm inner}$ & 12.9 & 14.5 & 14.4 & 15.5 & 13.4 & 13.2
& 12.3 & 12.9 & 12.9 & 12.9 & 12.9 & 12.8 & 12.9 & 13.0 \\ 
\end{tabular}}
\end{table*}

\begin{table}[h]
\caption{\label{tab:AshNucB} Same as Table~\ref{tab:AshNucA}, but
for the remaining rate variations.}
\resizebox{0.75\columnwidth}{!}{\begin{tabular}{c|ccccc}
A	&	15	&	16	&	17	&	18	&	19	\\
\hline											
4	&	9.97E-04	&	1.74E-03	&	9.99E-04	&	1.01E-03	&	1.13E-03	\\
12	&	4.77E-03	&	4.13E-03	&	4.94E-03	&	5.04E-03	&	5.21E-03	\\
13	&	7.15E-12	&	3.76E-12	&	1.60E-14	&	9.83E-14	&	5.40E-15	\\
16	&	3.07E-05	&	2.35E-05	&	2.81E-05	&	2.99E-05	&	2.96E-05	\\
19	&	5.13E-13	&	1.02E-13	&	1.55E-14	&	3.15E-17	&	3.18E-13	\\
20	&	7.36E-05	&	6.32E-05	&	4.47E-05	&	3.88E-05	&	4.07E-05	\\
21	&	1.09E-08	&	5.29E-09	&	6.16E-10	&	1.48E-10	&	2.85E-09	\\
22	&	1.86E-08	&	5.94E-09	&	9.83E-10	&	9.24E-13	&	4.46E-10	\\
23	&	3.20E-08	&	6.98E-08	&	1.73E-08	&	1.22E-09	&	2.28E-07	\\
24	&	5.23E-04	&	3.61E-04	&	4.47E-04	&	4.09E-04	&	4.55E-04	\\
25	&	2.83E-06	&	4.54E-06	&	1.77E-06	&	4.33E-07	&	1.31E-06	\\
26	&	2.27E-05	&	5.72E-05	&	2.02E-05	&	6.94E-07	&	4.94E-06	\\
27	&	3.60E-05	&	1.12E-04	&	4.13E-05	&	1.51E-06	&	6.88E-06	\\
28	&	7.31E-02	&	6.31E-02	&	7.36E-02	&	7.42E-02	&	7.07E-02	\\
29	&	3.22E-04	&	2.34E-04	&	3.94E-04	&	3.05E-04	&	2.71E-04	\\
30	&	1.90E-03	&	1.20E-03	&	1.92E-03	&	1.80E-03	&	2.08E-03	\\
31	&	3.35E-03	&	2.75E-03	&	3.38E-03	&	3.36E-03	&	3.58E-03	\\
32	&	4.60E-02	&	5.25E-02	&	4.44E-02	&	4.44E-02	&	4.62E-02	\\
33	&	6.17E-03	&	6.15E-03	&	6.14E-03	&	6.08E-03	&	6.62E-03	\\
34	&	1.25E-02	&	1.14E-02	&	1.26E-02	&	1.25E-02	&	1.39E-02	\\
35	&	4.48E-03	&	4.36E-03	&	4.54E-03	&	4.39E-03	&	4.88E-03	\\
36	&	4.22E-03	&	5.25E-03	&	4.18E-03	&	3.99E-03	&	4.40E-03	\\
37	&	7.12E-04	&	8.71E-04	&	7.02E-04	&	6.89E-04	&	7.54E-04	\\
38	&	4.12E-03	&	4.21E-03	&	4.19E-03	&	4.04E-03	&	4.64E-03	\\
39	&	3.99E-03	&	4.18E-03	&	3.99E-03	&	6.07E-03	&	4.38E-03	\\
40	&	2.15E-03	&	2.03E-03	&	2.17E-03	&	2.63E-03	&	2.11E-03	\\
41	&	5.28E-05	&	5.60E-05	&	5.46E-05	&	5.58E-05	&	4.91E-05	\\
42	&	6.22E-04	&	6.32E-04	&	6.25E-04	&	6.81E-04	&	7.10E-04	\\
43	&	3.09E-04	&	3.21E-04	&	2.95E-04	&	3.24E-04	&	3.26E-04	\\
44	&	4.67E-04	&	4.58E-04	&	4.71E-04	&	4.93E-04	&	4.52E-04	\\
45	&	4.16E-04	&	4.33E-04	&	4.24E-04	&	4.41E-04	&	4.33E-04	\\
46	&	1.13E-03	&	1.14E-03	&	1.13E-03	&	1.07E-03	&	1.29E-03	\\
47	&	4.12E-04	&	3.90E-04	&	3.91E-04	&	3.97E-04	&	4.07E-04	\\
48	&	7.81E-04	&	7.63E-04	&	7.96E-04	&	7.95E-04	&	7.88E-03	\\
49	&	8.89E-04	&	8.36E-04	&	8.92E-04	&	8.72E-04	&	9.38E-04	\\
50	&	1.23E-03	&	1.23E-03	&	1.22E-03	&	1.16E-03	&	1.09E-03	\\
51	&	1.86E-03	&	1.89E-03	&	1.84E-03	&	1.74E-03	&	1.48E-03	\\
52	&	3.89E-03	&	3.80E-03	&	3.94E-03	&	4.00E-03	&	2.06E-03	\\
53	&	1.32E-03	&	1.28E-03	&	1.33E-03	&	1.34E-03	&	8.05E-04	\\
54	&	1.39E-03	&	1.34E-03	&	1.38E-03	&	1.32E-03	&	1.09E-03	\\
55	&	3.43E-03	&	3.36E-03	&	3.43E-03	&	3.27E-03	&	2.81E-03	\\
56	&	1.36E-02	&	1.10E-02	&	8.81E-03	&	8.15E-03	&	8.79E-03	\\
57	&	4.15E-03	&	3.26E-03	&	3.63E-03	&	3.63E-03	&	2.69E-03	\\
58	&	1.62E-03	&	1.38E-03	&	1.57E-03	&	1.58E-03	&	1.25E-03	\\
59	&	3.82E-03	&	3.87E-03	&	4.02E-03	&	3.95E-03	&	3.44E-03	\\
60	&	1.14E-01	&	1.48E-01	&	1.17E-01	&	1.16E-01	&	1.13E-01	\\
61	&	1.42E-02	&	1.30E-02	&	1.41E-02	&	1.42E-02	&	1.37E-02	\\
62	&	8.49E-03	&	7.34E-03	&	8.36E-03	&	8.47E-03	&	8.18E-03	\\
63	&	1.73E-02	&	1.53E-02	&	1.71E-02	&	1.71E-02	&	1.68E-02	\\
64	&	3.43E-01	&	3.62E-01	&	3.46E-01	&	3.44E-01	&	3.36E-01	\\
65	&	2.58E-02	&	2.21E-02	&	2.58E-02	&	2.59E-02	&	2.66E-02	\\
66	&	1.33E-02	&	1.09E-02	&	1.33E-02	&	1.33E-02	&	1.38E-02	\\
67	&	1.43E-02	&	1.18E-02	&	1.44E-02	&	1.47E-02	&	1.50E-02	\\
68	&	1.11E-01	&	1.04E-01	&	1.12E-01	&	1.11E-01	&	1.14E-01	\\
69	&	1.93E-02	&	1.46E-02	&	1.93E-02	&	1.93E-02	&	2.03E-02	\\
70	&	6.56E-03	&	4.70E-03	&	6.44E-03	&	6.51E-03	&	7.06E-03	\\
71	&	6.62E-03	&	4.81E-03	&	6.53E-03	&	6.64E-03	&	7.15E-03	\\
72	&	3.19E-02	&	2.83E-02	&	3.20E-02	&	3.19E-02	&	3.32E-02	\\
73	&	1.09E-02	&	7.75E-03	&	1.08E-02	&	1.08E-02	&	1.14E-02	\\
74	&	4.08E-03	&	2.73E-03	&	3.99E-03	&	4.05E-03	&	4.43E-03	\\
75	&	5.13E-03	&	3.40E-03	&	5.03E-03	&	5.08E-03	&	5.54E-03	\\
76	&	9.69E-03	&	8.23E-03	&	9.58E-03	&	9.67E-03	&	9.98E-03	\\
77	&	4.69E-03	&	3.43E-03	&	4.67E-03	&	4.72E-03	&	4.80E-03	\\
78	&	3.93E-03	&	2.64E-03	&	3.86E-03	&	3.92E-03	&	4.11E-03	\\
79	&	3.71E-03	&	2.50E-03	&	3.64E-03	&	3.71E-03	&	3.82E-03	\\
80	&	3.74E-03	&	3.19E-03	&	3.65E-03	&	3.75E-03	&	3.76E-03	\\
81	&	2.38E-03	&	1.86E-03	&	2.35E-03	&	2.42E-03	&	2.35E-03	\\
82	&	3.48E-03	&	2.56E-03	&	3.42E-03	&	3.56E-03	&	3.43E-03	\\
83	&	2.06E-03	&	1.58E-03	&	2.01E-03	&	2.12E-03	&	2.02E-03	\\
84	&	1.68E-03	&	1.43E-03	&	1.62E-03	&	1.72E-03	&	1.65E-03	\\
85	&	1.90E-03	&	1.75E-03	&	1.84E-03	&	1.97E-03	&	1.83E-03	\\
86	&	1.60E-03	&	1.44E-03	&	1.56E-03	&	1.68E-03	&	1.53E-03	\\
87	&	1.54E-03	&	1.51E-03	&	1.49E-03	&	1.61E-03	&	1.46E-03	\\
88	&	6.95E-04	&	7.59E-04	&	6.59E-04	&	7.01E-04	&	6.51E-04	\\
89	&	1.05E-03	&	1.22E-03	&	1.00E-03	&	1.03E-03	&	9.53E-04	\\
90	&	3.17E-04	&	3.74E-04	&	2.97E-04	&	2.98E-04	&	2.77E-04	\\
91	&	2.76E-04	&	3.60E-04	&	2.64E-04	&	2.58E-04	&	2.46E-04	\\
92	&	2.03E-04	&	2.62E-04	&	1.91E-04	&	1.84E-04	&	1.76E-04	\\
93	&	1.85E-04	&	2.57E-04	&	1.71E-04	&	1.62E-04	&	1.60E-04	\\
94	&	4.91E-04	&	7.85E-04	&	4.53E-04	&	4.02E-04	&	4.28E-04	\\
95	&	7.35E-05	&	8.49E-05	&	7.69E-05	&	7.04E-05	&	6.35E-05	\\
96	&	6.17E-05	&	1.02E-04	&	5.74E-05	&	4.88E-05	&	5.39E-05	\\
97	&	7.32E-05	&	1.30E-04	&	6.83E-05	&	5.53E-05	&	6.41E-05	\\
98	&	7.09E-05	&	1.46E-04	&	6.75E-05	&	4.97E-05	&	6.33E-05	\\
99	&	4.51E-05	&	9.77E-05	&	4.58E-05	&	3.13E-05	&	4.23E-05	\\
100	&	2.32E-05	&	5.53E-05	&	2.49E-05	&	1.53E-05	&	2.29E-05	\\
101	&	1.21E-05	&	3.56E-05	&	1.36E-05	&	7.14E-06	&	1.30E-05	\\
102	&	4.53E-06	&	1.59E-05	&	5.49E-06	&	2.43E-06	&	5.32E-06	\\
103	&	1.44E-06	&	5.95E-06	&	1.91E-06	&	6.94E-07	&	1.82E-06	\\
104	&	2.58E-07	&	1.24E-06	&	3.67E-07	&	1.18E-07	&	3.33E-07	\\
105	&	1.14E-08	&	5.91E-08	&	1.72E-08	&	4.58E-09	&	1.38E-08	\\
\hline
$Q_{\rm imp}^{\rm inner}$ & 13.0 & 13.7 & 12.9 & 12.9 & 13.1 \\
\end{tabular}}
\end{table}

\section{Influence on Nucleosynthesis}
\label{sec:ashes}
Upon reproducing an observed burst light curve, it is natural to
surmise that the nucleosynthesis for a particular source has also
been reproduced. This could provide constraints on the composition
and thermal structure of the outer layers of the neutron star, which
would be of use when modeling other accreting neutron star
observables~\citep{Deib16,Meis17}. We therefore also investigate the
impact of nuclear reaction rate uncertainties on the X-ray burst ash
composition and compare this to the impact of modified
astrophysical conditions.

\subsection{Surface Abundances}
\label{ssec:surfaceash}

Ash abundances were extracted after the last burst in the burst
sequence, i.e. after $\sim10-20$ bursts, for a consistent connection
to the averaged light curve. However, we note that we observe the
burst abundances converge after $\sim$4~bursts. Mass fractions
$X(A)$ for species of nuclear mass $A$ were calculated by averaging
over envelope layers that no longer experienced hydrogen or helium
burning, following the approach of \citet{Cybu16,Meis17}.

Figure~\ref{fig:AshAstro} and Table~\ref{tab:AshAstro} show the ash abundance distributions
corresponding to the calculated light curves of
Figure~\ref{fig:LCastro}. Qualitatively, decreasing $\dot{M}$,
decreasing $X(H)$, and increasing $Q_{\rm b}$ all reduce the extent
of the ash distribution to lower-$A$ material. 

For the lowest
$\dot{M}$ modeled here, $X(A)$ rapidly falls off after $A=60$, with
steadily decreasing resurgences for the $A=60+4n$ sequence, where $n$
is an integer. The $X(A)$ distribution steadily fills in up to
$A\sim100$ for increasing $\dot{M}$, washing-out the $A=4n$
signature for $A>60$. This increase comes at the expense of $X(A)$
for lower-$A$ nuclides. It is notable that lower $\dot{M}$ results in a
larger $X(12)$, as current models have difficulty explaining the
amount of $^{12}{\rm C}$ required to ignite X-ray
superbursts. Though all models fall far short of the  $X(^{12}{\rm
C})\gtrsim20$\% required to reconcile superburst models with
observations~\citep{Cumm06}. These findings are consistent with
lower $\dot{M}$ resulting in more helium-rich bursts, therefore
favoring $\alpha$-capture reactions and as a result nuclei with
$A=4n$. 

Unsurprisingly, directly lowering $X(H)$ of the accreted
material leads to similar behavior in the ash composition. For
$X(H)=0.75$, which is just above the solar surface value, a
significant enhancement for $A\gtrsim80$ occurs, including a three
orders of magnitude increase for $X(100)$.

Increasing $Q_{\rm b}$ over the range explored here has a more
modest impact than the changes we observe for our $\dot{M}$ and
$X(H)$ variations. The reduction in high-$A$ abundances is likely
due to the absence of fuel left to burn late in the burst evolution
because of the reduced fuel build-up time between bursts. We note
that increased $Q_{\rm b}$ leads to increased $X(12)$, in qualitative
agreement with \citet{Reic17}, though calculation details are
too different for a more direct comparison.

The different extent of the ash distributions in $A$ can be seen
when examining the flow of nucleosynthesis in the $rp$-process. The
integrated flow for converting isotope $i$ to isotope $j$ is
$\mathcal{F}_{ij}=\int\left(\dot{Y}_{i\rightarrow
j}-\dot{Y}_{j\rightarrow i}\right)dt$ where the integral runs over
the full burst time and $\dot{Y}_{i\rightarrow j}$ is the rate that the
abundance of isotope $i$ is depleted by reactions converting $i$ to
$j$. $\mathcal{F}_{ij}$ for the models best reproducing GS 1826-24,
with $X(H)$ increased to 0.75, and $\dot{M}$ lowered to
0.07~$\dot{M}_{\rm E}$ are shown in Figure~\ref{fig:flow} for
demonstration purposes. The different flows highlight the fact that
reaction rate sensitivities may be quite different for different
astrophysical conditions.

Examples for the impact of nuclear reaction rate variations on the
surface ash composition are shown in Figure~\ref{fig:AshRxn}, where
the composition for all rate variations of Section~\ref{sec:rates}
are given in Tables~\ref{tab:AshNucA} and \ref{tab:AshNucB}.
Comparing to Figure~\ref{fig:AshAstro}, it is apparent that rate
variations have a comparable impact to modest changes in the
accretion conditions. Nonetheless, for some reactions 
orders of magnitude changes in
$X(A)$ can be observed at the high and low-$A$ edges of the
composition distribution and, for several other reactions, local
changes near $A$ of the reactant isotope can be observed. These
local changes can often far exceed the variation in $X(A)$ due to
varied accretion conditions, e.g.
$A=59$ for $^{59}{\rm Cu}(p,\gamma)/100$.

\subsection{Crust Temperature and Composition}
\label{ssec:crust}
Ashes produced on the accreting neutron star surface are buried by
subsequent accretion to higher densities. There they are steadily
modified by nuclear reactions, resulting in some compositional
structure and driving local heat
sources and heat sinks~\citep{Lau18}. These modifications to the
compositional and thermal structure ultimately influence the
ignition of X-ray superbursts and the light curves of cooling
transient neutron stars~\citep{Deib16,Meis17}.
Here we focus on heat sources and sinks which have previously been
identified as significant and the composition of the neutron star
inner crust, whose connection to the surface burning ashes has been
recently calculated using detailed reaction network calculations.

Heating from electron-capture is
strongest for even-$A$ nuclides with a large odd-even mass
staggering, since for these cases a significant fraction of
the second electron-capture in a two electron-capture sequence 
can proceed through excited states that then
deposit a large amount of energy through
$\gamma$-deexcitation. Among these, $^{66}{\rm
Ni}$$\rightarrow$$^{66}\rm{Fe}$, $^{66}\rm{Fe}$$\rightarrow$$^{66}{\rm
Cr}$, and $^{64}\rm {Cr}$$\rightarrow$$^{64}\rm{Ti}$ stand
out~\citep{Gupt07}. We find nuclear reaction rate uncertainties could
allow for at most a factor of $\sim2$ change in $X(64)$ and $X(66)$.
$X(64)$ shows a similar level of stability across the range of
astrophysical conditions modeled here, though $X(66)$ can be
substantially reduced from the possible percent level for
particularly helium-rich bursts.

Local heat sinks can be formed in the neutron star outer layers by
cycling between electron captures and $\beta^{-}$-decays, which
predominantly occurs for odd-$A$ nuclides in regions of deformation
with particularly small odd-even mass
staggering~\citep{Scha14,Meis15}. The potential for this urca cooling
process has been identified for $A=29,31,33,55,57,59,63,65,103$, and
$105$, where $X(A)\sim5\times10^{-3}$ or greater are required for significant
cooling~\citep{Meis17}. We find that $^{15}{\rm O}(\alpha,\gamma)/10$
enhances $X(29)$ and $X(31)$ to this level. $X(33)$, $X(55)$, and
$X(57)$ are always at or
near the threshold of significant cooling for all rate variations.
$X(59)$ is generally below the threshold, though is boosted to
$X(A)\sim5\times10^{-2}$ by $^{59}{\rm Cu}(p,\gamma)/100$. $X(63)$
and $X(65)$ are at the percent level for all rate variations, though
it is notable that $X(63)$ is doubled from the nominal value by
$^{61}{\rm Ga}(p,\gamma)/100$, likely because this variation
effectively stalls material within the ZnGa cycle. $X(103)$ and
$X(105)$ are sensitive to a number of the nuclear reaction rate
variations explored here, though they are orders of magnitude too
small to result in significant urca cooling for each of these
calculations.  

In the neutron star inner crust the thermal conductivity is
dominated by electron-ion impurity scattering. As such, the inner
crust impurity is of particular interest. This is quantified using
the variance of the charge distribution, which is known as the
impurity parameter $Q_{\rm imp}$.
$Q_{\rm imp}=\frac{1}{n_{\rm
ion}}\Sigma_{j}n_{j}\left(Z_{j}-\langle Z\rangle\right)^{2}$, where
$n_{\rm ion}$ is the ion number density, $n_{j}$ is the number
density of species with atomic number $Z_{j}$, and $\langle
Z\rangle$ is the average atomic number of the composition. 
We cannot directly infer $Q_{\rm imp}$ from the surface burning
ashes, since crust reactions drastically modify the composition as
it is buried to inner crust depths. However, \citet{Lau18} found a
correlation between the $A$ of a nucleus at the surface and its
corresponding equilibrium isotope in the crust. At the inner crust,
material becomes concentrated into the $N=28$, 50, and 82
shell closures; for the finite range droplet mass model
(FRDM)~\citep{Moll95}, these are $^{40}{\rm Mg}$, $^{70}{\rm Ca}$,
and $^{116}{\rm Se}$. The surface to inner crust mapping they found
was that $A<28$ and $A=29-55$ are funneled into $N=28$, $A=56-60$ are
split between $N=28$ and $N=50$, $A=61-101$ ultimately go to $N=50$,
and $A=102-105$ are split between $N=50$ and $N=82$. For the cases
where $X(A)$ are split between two magic numbers, \citet{Lau18} note
that the exact split depends on the neutron abundance and competing
reaction rates. 

We calculate the inner crust impurity $Q_{\rm imp}^{\rm inner}$
resulting from our surface abundance distributions
using the mapping just described, where we use the crude
approximation that splitting between two magic numbers is half and
half. These results are in the last row of
Tables~\ref{tab:AshAstro}--\ref{tab:AshNucB}. To be clear, a precise calculation would require use of a crust reaction
network for varied astrophysical conditions and nuclear mass models.
Nonetheless, these calculations are informative as a first
approximation to the impact of nuclear reaction rate uncertainties
and accretion conditions on the inner crust impurity. We find
$10\lesssim Q_{\rm imp}^{\rm inner}\lesssim16$ for all modeled
conditions. The impurity is smallest for the most helium-rich
conditions, increasing for moderately helium-rich bursts, and
decreasing again for the most hydrogen-rich bursts. This is because
the ashes from helium-rich bursts have a large fraction of nuclides that become
$N=28$ in the inner crust, whereas  more hydrogen-rich bursts result
in a larger fraction of $N=50$. $Q_{\rm
imp}^{\rm inner}$ is largest in the intermediate region. The nuclear
reaction rate variations explored here can have a comparable impact on
$Q_{\rm imp}^{\rm inner}$. Specifically, $\gtrsim10$\% enhancements
are observed for $^{59}{\rm Cu}(p,\alpha)\times100$,
$^{59}\rm{Cu}(p,\gamma)/100$, and $^{61}{\rm Ga}(p,\gamma)/100$.
Each of these reaction rate variations boost $X(A)$ in the split
region $A=56-60$ and so the impact noted here may be an artifact of
the adopted splitting ratio.

\section{Influence on the Neutron Star Mass-Radius
Ratio Derived from
Model-Observation Comparisons}
\label{sec:compare}

In Section~\ref{ssec:relative}, it was shown that relatively tight
constraints can be placed on the accretion conditions for GS 1826-24
based on reproducing the observed light curve shape and recurrence
time over a range of accretion rates. This restricts the number of
free parameters used to reproduce the observed light curve, which
therefore reduces the phase space for remaining free parameters.
Namely, we can now extract constraints for $d\xi_{\rm b}^{1/2}$ and
$1+z$. This procedure was described by \citet{Meis18a}, but is
briefly repeated here for completeness. We then demonstrate how
these constraints are sensitive to uncertainties in nuclear reaction
rates and how this sensitivity translates into an uncertainty on the
extracted neutron star mass-radius ratio. Note, the $1+z$
determination may be systematically altered by astrophysical
effects, such as the flame spreading on the neutron star surface, as
pointed-out by \citet{Zamf12}. As such \emph{one should focus on the
relative change of $1+z$ with modified nuclear reaction rates,
rather than on the absolute determination of $1+z$}.

\subsection{Model-Observation Comparison of the Light Curve Shape}
Observed light curve data for GS 1826-24 come from the analysis of
the data in the Multi-Instrument Burst Archive
(MINBAR)\footnote{https://burst.sci.monash.edu/minbar/}
described in \citet{Gall17}. They took individual burst light curves
observed over a bursting epoch with the Rossi X-ray Timing Explorer~\citep{Gall04,Gall08},
aligned them in time, and obtained an average flux at each time,
which is adopted as the observed light curve.

The simulated light curve is altered for an assumed
$d\xi_{\rm b}^{1/2}$ and $1+z$
to create
$F(\tilde{t})=L(\tilde{t})/(4\pi\xi_{\rm{b}}(1+z)d^{2})$, where $F$ is the
flux, $L$ is the simulated luminosity, and $\tilde{t}=(1+z)t$ is the
redshifted time~\citep{Gall17}. 
The neutron star mass and radius adopted for the
simulations correspond to $(1+z)=1.26$, so in this sense choosing other redshifts
is inconsistent. However, in practice burst ignition properties are
insensitive to modest changes in the neutron star mass $M_{\rm{NS}}$
and radius $R_{\rm{NS}}$, and therefore $(1+z)$~\citep{Ayas82,Zamf12}.
Finally, to compare to the observed light curve, a time shift $\delta
t$ is applied to the simulated light curve to obtain the minimum
$\chi^{2}_{\rm red}$ for a given 
combination of $d\xi_{\rm b}^{1/2}$ and $1+z$. 

We calculate $\chi^{2}_{\rm red}$ using the 50~s of the light
curve following thermonuclear runaway\footnote{The absolute $1+z$ is sensitive to this range, for
instance shifting downward by $\sim0.05$ when the fit-range is
extended to cover the 150~s following thermonuclear runaway.
However,
the relative effect of the reaction rate variations is not significantly changed.}
over the phase space $4.0\leq d\xi_{\rm
b}^{1/2}\leq8.1$~kpc in steps of 0.1~kpc, $1.16\leq(1+z)\leq1.84$ in steps
of 0.01, and $0.2\leq\delta t\leq0.8$~s in steps of 0.1~s. See
Figure 3 of \citet{Meis18a} for an example of the  $\chi^{2}_{\rm
red}$ surface for the baseline reaction rate library. We assume our
uncertainties for the parameters $d\xi_{\rm b}^{1/2}$ and $1+z$ are Gaussian
in order to convert between
$\Delta\chi^{2}$ from the minimum $\chi^{2}$ and a confidence interval so that we can more
easily compare $\chi_{\rm red}^{2}$ surfaces obtained for different
model calculations~\citep{Pres92}.

\begin{figure}[t]
\begin{center}
\includegraphics[width=1.0\columnwidth,angle=0]{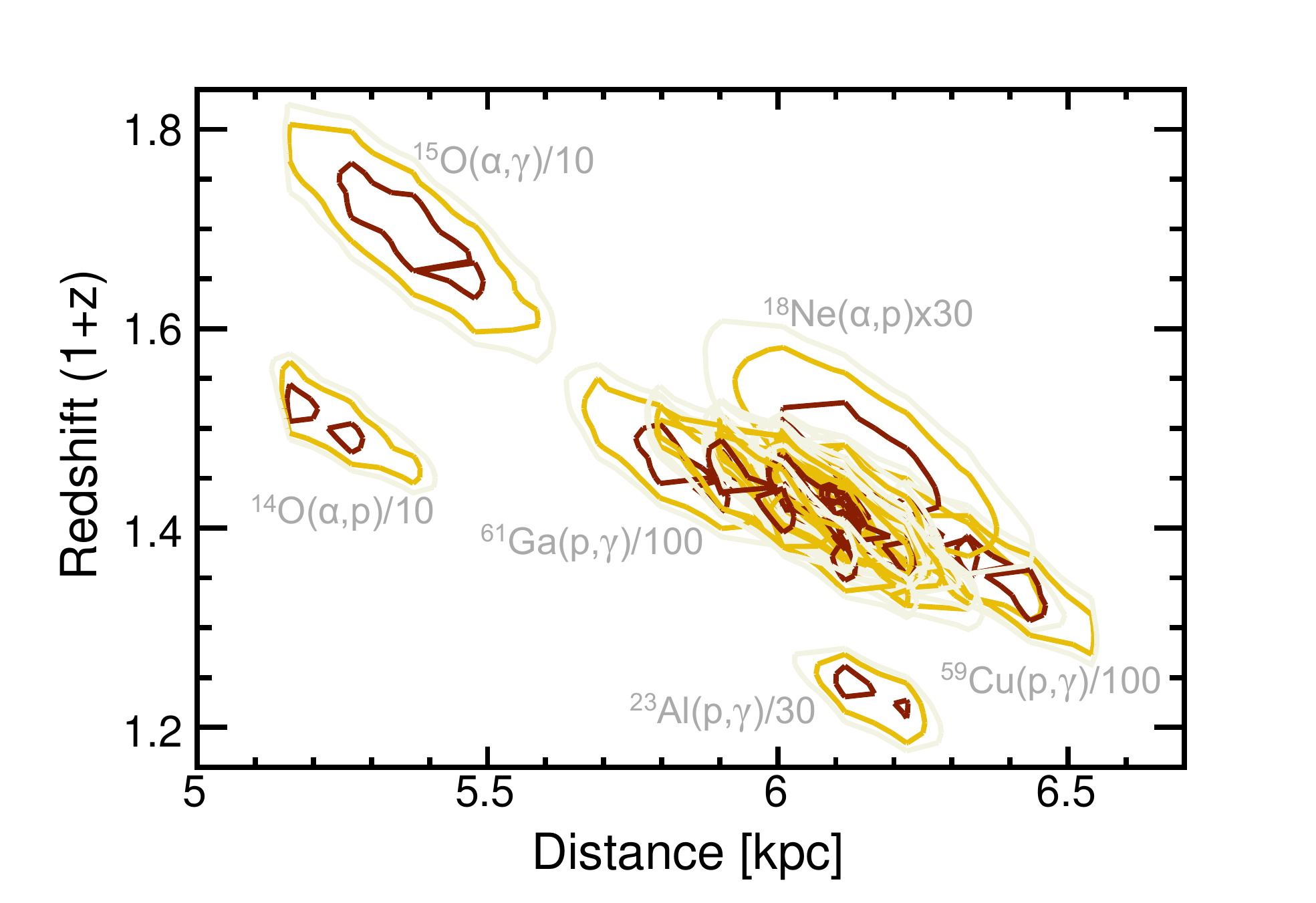}
\caption{68\% (red lines), 95\% (yellow lines), and 99\% (gray
lines) confidence intervals for the distance-redshift determination
performed comparing light curves shown in Figure~\ref{fig:LCall} to
the GS 1826-24 2007 bursting epoch.
Cases significantly deviating from the result for the baseline
calculation are labeled.
\label{fig:Confidence}}
\end{center}
\end{figure}

\subsection{Impact of Nuclear Reaction Rates on the Extracted
Distance and Redshift}
The technique of the previous subsection was applied using each of
the model calculations shown in Figure~\ref{fig:LCall}. The
resulting confidence intervals for $d\xi_{\rm b}^{1/2}$ and $1+z$ are 
shown in Figure~\ref{fig:Confidence}. We identify six reaction rate
variations
which result in significantly different $d\xi_{\rm b}^{1/2}$ and/or
$1+z$ compared to the baseline calculation: $^{15}{\rm
O}(\alpha,\gamma)/10$, $^{14}{\rm O}(\alpha,p)/10$, $^{18}{\rm
Ne}(\alpha,p)\times30$, $^{23}{\rm Al}(p,\gamma)/30$, $^{59}{\rm
Cu}(p,\gamma)/100$, and $^{61}{\rm Ga}(p,\gamma)/100$. That these
are the influential rates is unsurprising, due to their impact on
$t_{\rm rise}$, $\mathcal{C}$, \cancel{E}, and the peak flux. 

Only $^{15}{\rm O}(\alpha,\gamma)/10$ results in a significantly
increased $1+z$.
We can therefore determine that an increase of
$\mathcal{C}$ of $\sim75$\% or larger is consequential for the
redshift determination. $^{23}{\rm Al}(p,\gamma)/30$ results in a
significant reduction in $1+z$ from the baseline result, which may
be a combination of the increase in $t_{\rm rise}$ and decrease in
$\mathcal{C}$. Nonetheless, the absence of a significantly reduced
$1+z$ for $^{18}{\rm Ne}(\alpha,p)\times30$ indicates that
$\mathcal{C}$ reductions of 25\% or less are not consequential for
the redshift determination. The influence of $\cancel{E}$ on the
best-fit $1+z$ is apparent from $^{59}{\rm Cu}(p,\gamma)/100$, for
which an \cancel{E} increase results in a reduced $1+z$, and
$^{61}{\rm Ga}(p,\gamma)$, for which a \cancel{E} decrease results
in an increased $1+z$. Since $^{22}{\rm Mg}(\alpha,p)/10$ does not
stand-out in this phase space, this sets a rough limit for an
increase of \cancel{E} that is consequential. 

The deviations of
$d\xi_{\rm b}^{1/2}$ from the baseline result stem from the altered
peak luminosity, with the exception of $^{15}{\rm
O}(\alpha,\gamma)/10$ for which the large increase in $\mathcal{C}$
also plays a role. From Figure~\ref{fig:LCall}, it is clear that relatively
modest alterations of the peak luminosity lead to quite different
distance determinations. For instance, the $\sim15$\% decrease in
peak luminosity for the calculation with $^{14}{\rm O}(\alpha,p)/10$
reduces the best-fit distance by roughly the same amount.

\begin{figure}[t]
\begin{center}
\includegraphics[width=1.0\columnwidth,angle=0]{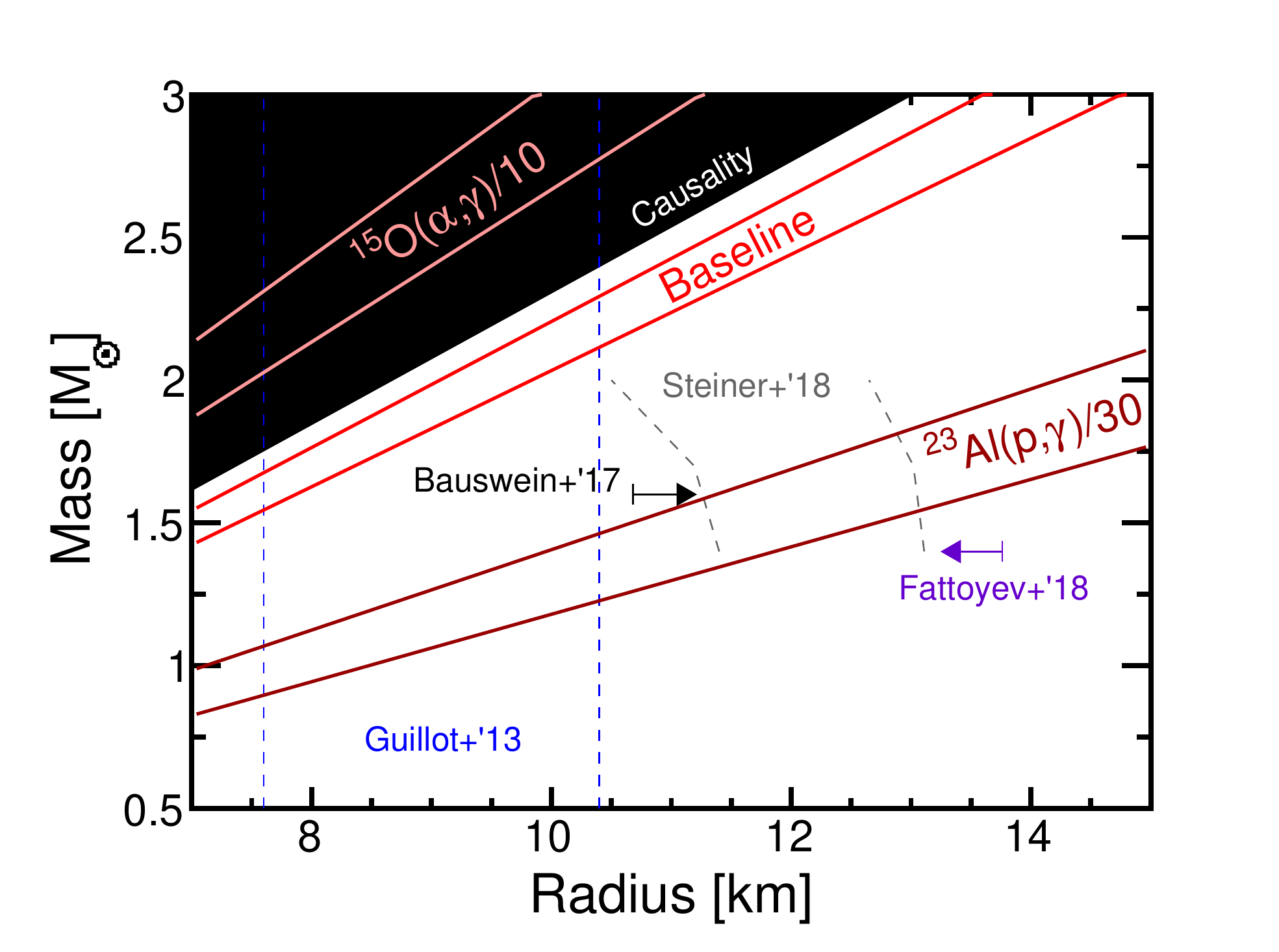}
\caption{$M_{\rm NS}/R_{\rm NS}$ bands calculated from the $1+z$
68\% confidence intervals shown in Figure~\ref{fig:Confidence} for
the baseline calculation and the two reaction rate variations
resulting in the extremes for $1+z$. Constraints from other
astrophysical observables~\citep{Guil13,Baus17,Fatt18,Stei18} are also shown and the unphysical
region excluded by causality~\citep{Kora97} is shown in black.
\label{fig:MRdiagram}}
\end{center}
\end{figure}

\subsection{Implications for Neutron Star Mass-Radius Ratio
Constraints}
The constraint for $1+z$ can directly be translated into a
constraint on the neutron star mass and radius using the
relationships described in the appendix of \citet{Lamp16}. The
general relativistic neutron star mass and radius are
related to the surface gravitational redshift by
$1+z=1/\sqrt{1-2GM_{\rm GR}/\left(R_{\rm
GR}c^{2}\right)}$, where $c$ is the speed of light and $G$ is the
gravitational constant. For the choice that the Newtonian mass is equal
to the general relativistic mass $M_{\rm NS}=M_{\rm GR}$, then
$R_{\rm GR}=\sqrt{(1+z)}R_{\rm NS}$. Therefore an uncertainty in $1+z$
can be converted into a band on the neutron star mass-radius
diagram. This is shown for three calculations in
Figure~\ref{fig:MRdiagram}. It is clear the uncertainties in nuclear
reaction rates prevent the use of model-observation comparisons to
constrain $M_{\rm NS}/R_{\rm NS}$ to a useful degree. In particular,
for the astrophysical conditions adopted here to reproduce the
		  observed characteristics of GS 1826-24,
improved uncertainties for $^{15}{\rm
O}(\alpha,\gamma)$ and $^{23}{\rm Al}(p,\gamma)$ are essential.

Eliminating the nuclear physics uncertainties that alter the
determination of $1+z$ will open up the possibility to employ this
light curve matching for a new constraint on $M_{\rm NS}/R_{\rm
NS}$. Of course, the uncertainty contributions from astrophysical
phenomena like flame spreading, e.g. the impact on $\mathcal{C}$ from
polar versus equatorial ignition~\citep{Maur08}, need to be explored
further. We leave this for future work. 
If we take our present constraints at face value, we could
possibly constrain nuclear reaction rates. Namely,
Figure~\ref{fig:MRdiagram} shows that  the $^{15}{\rm
O}(\alpha,\gamma)$ reaction rate cannot be a factor of 10 lower than
the median rate of \citet{Davi11}\footnote{This conclusion was reached by
\citet{Meis18a} based on reproducing $\mathcal{C}$ of the observed
light curve.}.

If one adopts a particular dense matter equation of state, then our
constraint can be used to determine $M_{\rm NS}$. Alternatively, we
could assume a stiff equation of state and use other astrophysical
constraints for $R_{\rm NS}$, or perhaps a photospheric radius expansion
burst for the same bursting source, to determine $M_{\rm NS}$. Either
approach would provide additional data for the $M_{\rm NS}$ mass
distribution, which is of interest for understanding the fates of massive stars.

\section{Conclusions}
\label{sec:discuss}
To summarize, we assessed the influence of nuclear reaction rate
variations on the results of multizone X-ray burst model
calculations performed with the code {\tt MESA} with respect to the
influence of modified astrophysical conditions.  Considering the 19
most influential reaction rate variations for the X-ray burst light curve
identified by \citet{Cybu16} and using the astrophysical conditions
that \citet{Meis18a} found best reproduce X-ray bursts from the GS
1826-24 years 1998, 2000, and 2007 bursting epochs, we assessed the
impact of these rate variations on properties of the X-ray burst
light curve and X-ray burst ashes. 

We found that notable impacts on
features of the X-ray burst light curve occur for the reaction rate
variations $^{14}{\rm O}(\alpha,p)/10$, $^{15}{\rm
O}(\alpha,\gamma)/10$ $^{18}{\rm Ne}(\alpha,p)\times30$, $^{22}{\rm
Mg}(\alpha,p)/10$, $^{24}{\rm Mg}(\alpha,\gamma)\times10$,
$^{23}{\rm Al}(p,\gamma)/30$, $^{59}{\rm
Cu}(p,\gamma)/100$, 
and $^{61}{\rm Ga}(p,\gamma)/100$.  However, none of the impacts are
large enough to significantly change the conclusions about the
accretion conditions for GS 1826-24 that were inferred by
\citet{Meis18a}. 

We found that heating from electron-capture on the buried ashes in
the neutron star crust is largely insensitive to the nuclear
reaction rate variations we investigated and was only significantly
reduced for particularly helium-rich bursts. On the other hand,
cooling from urca cycles involving buried ashes could be enhanced by
the rate variations $^{15}{\rm O}(\alpha,\gamma)/10$, $^{59}{\rm
Cu}(p,\gamma)/100$, and $^{61}{\rm Ga}(p,\gamma)/100$. Using a
heuristic for mapping between surface abundances and the inner crust
impurity, we determined that $^{59}{\rm Cu}(p,\alpha)\times100$,
$^{59}{\rm Cu}(p,\gamma)/100$, and $^{61}{\rm Ga}(p,\gamma)/100$ can
likely influence $Q_{\rm imp}^{\rm inner}$ on a similar order as
significant changes in accretion conditions; however, the overall
changes are still modest.

Given the narrow range of astrophysical conditions that can
reproduce the observed features of GS 1826-24, we were able to also
determine constraints on the distance and surface gravitational
redshift of this source, with the caveat that significant
astrophysical uncertainties remain. Nonetheless, we demonstrate that
reaction rate uncertainties are an additional obstacle to obtaining
these constraints. Specifically, an influence on $d\xi_{\rm
b}^{1/2}$ and/or $1+z$ is observed for the rate variations $^{15}{\rm
O}(\alpha,\gamma)/10$, $^{14}{\rm O}(\alpha,p)/10$, $^{18}{\rm
Ne}(\alpha,p)\times30$, $^{23}{\rm Al}(p,\gamma)/30$, $^{59}{\rm
Cu}(p,\gamma)/100$, and $^{61}{\rm Ga}(p,\gamma)/100$. We identify
$^{15}{\rm O}(\alpha,\gamma)/10$ and $^{23}{\rm Al}(p,\gamma)/30$ as
particularly significant in this regard.

We stress that the reaction rates deemed influential may be
different for other astrophysical conditions. As such, reaction rate
sensitivity studies for a range of astrophysical conditions are
welcome. Studies identifying the sets of astrophysical
conditions that reproduce the light curve features observed for
other X-ray bursting sources are also desired. Finally, it would be
interesting to extend the present work to other nuclear physics
uncertainties that have been shown to impact the X-ray burst light
curve, e.g. nuclear masses~\citep{dSan14,Scha17}.

\begin{acknowledgments}
We thank the contributors to the
{\tt MESA} Marketplace
({\tt http://cococubed.asu.edu/mesa\_market/}) and the associated
forum, who enabled the
calculations presented here.
This work was supported by the U.S. Department of Energy
under grants DE-FG02-88ER40387 and DE-SC0019042 and benefited
from support by the National Science Foundation under grant
PHY-1430152 (Joint Institute for Nuclear Astrophysics--Center for the Evolution of the Elements).
\end{acknowledgments}

\bibliographystyle{apj}
\bibliography{RxnUncReferences}

\end{document}